\documentclass[prd,twocolumn,nofootinbib,showpacs,superscriptaddress]{revtex4}

\usepackage{amsfonts}
\usepackage{amsmath}
\usepackage{amssymb}
\usepackage{bm}
\usepackage{dcolumn}
\usepackage{epsfig}
\usepackage{graphicx}
\usepackage[latin1]{inputenc}
\usepackage{latexsym}
\usepackage{rotating}
\usepackage{hyperref}
\usepackage[usenames]{color}
\usepackage{float}
\usepackage{xspace} 
\usepackage{mathrsfs}
\usepackage{subfigure}
\usepackage{enumitem}
\usepackage{caption}
\usepackage{ulem}
\definecolor{darkgreen}{rgb}{0.2,0.7,0.2}

\newcommand\be{\begin{equation}}
\newcommand\ba{\begin{eqnarray}}
\newcommand\ee{\end{equation}}
\newcommand\ea{\end{eqnarray}}

\usepackage{setspace}

\begin{document}
\title{Towards Robust Gravitational Wave Detection with Pulsar Timing Arrays}

\author{Neil J. Cornish}
\affiliation{Department of Physics, Montana State University, Bozeman, MT 59717, USA.}

\author{Laura Sampson}
\affiliation{Department of Physics, Montana State University, Bozeman, MT 59717, USA.}

\date{\today}

\begin{abstract} 
Precision timing of highly stable milli-second pulsars is a promising technique for the detection of very low frequency sources of
gravitational waves. In any single pulsar, a stochastic gravitational wave signal appears as an additional source of timing noise that can be absorbed by the noise model, and so it is
only by considering the coherent response across a network of pulsars that the signal can be distinguished from other sources
of noise. In the limit where there are many gravitational wave sources in the sky, or many pulsars in the array,
the signals produce a unique tensor correlation pattern that depends only on the angular separation between each pulsar pair. It is
this distinct fingerprint that is used to search for gravitational waves using pulsar timing arrays. Here we consider how the
prospects for detection are diminished when the statistical isotropy of the timing array or the gravitational wave signal is broken
by having a finite number of pulsars and a finite number of sources.  We find the standard tensor-correlation analysis to
be remarkably robust, with a mild impact on detectability compared to the isotropic limit. Only when there are very few
sources and very few pulsars does the standard analysis begin to fail. Having established that the tensor correlations are a robust
signature for detection, we study the use of ``sky-scrambles'' to break the correlations as a way to increase confidence in a detection.
This approach is analogous to the  use of ``time-slides'' in the analysis of data from ground based interferometric detectors.
\end{abstract}
\maketitle
\allowdisplaybreaks[4]

\section{Introduction}

 With the steady addition of new pulsars to the arrays and improvements in the timing
sensitivity and analyses, pulsar timing is advancing rapidly as a technique for the detection of gravitational waves. Impressive new upper limits on the amplitude of power-law stochastic backgrounds are starting to challenge simple astrophysical models that
attribute the background to the gravitational wave driven evolution of a population of supermassive black hole binaries on quasi-circular orbits~\cite{Shannon25092015,2015MNRAS.453.2576L,2015arXiv150803024A}. These limits are dominated by
the timing residuals from one or two very low noise pulsars that have been observed for many years. A detection, on the other hand, will come from combining the data of
a very large number of moderately sensitive pulsars~\cite{Siemens:2013zla}. 

The key to making a detection, as opposed to setting an upper limit, is the unique correlation pattern that results when a gravitational wave signal passes through an array
of pulsars. In the limit where there are an infinite number of isotropically distributed sources~\cite{HellingsDowns} (and at least two pulsars) or an infinite number of
pulsars (and at least one source)~\cite{Cornish:2013aba}, the correlation in the timing residuals of two pulsars $a,b$
with an angular separation $\alpha_{ab}$ has the form
\begin{equation}\label{tensor}
H_{ab} = \frac{3 \, c_{ab}}{2} \, \ln c_{ab} - \frac{c_{ab}}{4}+\frac{1}{2}\left( 1 + \delta(\alpha_{ab})\right) \, ,
\end{equation}
where $c_{ab} = (1-\cos\alpha_{ab})/2$.  In any actual experiment, neither condition required to arrive at (\ref{tensor}) is met. There will only be a finite number of sources contributing to the signal, and a finite number of pulsars contributing to the timing array. This means that the standard correlation analysis will be sub-optimal~\cite{Cornish:2013aba}. Here we study the
impact that this has on the detectability of a stochastic background. We do this by comparing the detectability of the highly anisotropic signal formed from a finite number of black hole binaries to that of an idealized isotropic signal with the same average power
level. We investigate this as a function of the number of black holes and the number of pulsars. We find that the standard correlation analysis is remarkably robust, and only results in a small
loss in detection efficiency for realistic signals and array sizes. It is only in the limit of very few pulsars and very few sources that a significant loss of effectiveness occurs.

Having established the detection of a tensor correlation pattern as a robust signature of gravitational waves, we investigate the use of ``sky-scrambles'' to purposely break the signal correlations
in the data as a test of the analysis pipelines. If the evidence for a gravitational wave signal is largest for the true pulsar sky locations, and much smaller for any of the scrambled sky locations, then
we gain confidence in our models and analysis techniques. Each sky scramble produces a distinct correlation pattern, $C'_{ij}$, and we can define a measure of closeness based on the similarity
of the correlation patterns. As hoped, we find that the evidence for a gravitational wave signal is highest for the true pulsar locations, and lowest for scrambles that are most dissimilar to
the expected correlation pattern. Based on our simple measure of closeness, there are a limited number of independent sky scrambles, which limits the statistical power of the test. But we argue
against trying to use the test in a frequentist framework. Rather, sky-scrambles help to validate the noise and signal models used to compute the Bayesian evidence of a signal that,
like the cosmic microwave background, we only get to see once. 

Our work builds on several earlier studies~\cite{Cordes:2011vg,2011MNRAS.414.1777Y,2009PhRvD..79h4030A,2015PhRvD..91d4048C,2009PhRvD..79f2003F}, where various statistics are developed to
detect gravitational wave signals in pulsar timing data. Of particular relevance is the `optimal statistic'~\cite{2009PhRvD..79h4030A, 2014arXiv1410.8256C} for the detection of the stochastic background, which is essentially a measure of how important the cross-correlations between pulsars are for describing the signal. Both our method and these frequentist analyses investigate the detection of the tensor correlation pattern between pulsars, which we again emphasize is necessary for the unambiguous detection of gravitational waves. 

\section{The simulated astrophysical signal}

\begin{figure*}[htp]
\includegraphics[clip=true,angle=0,width=0.48\textwidth]{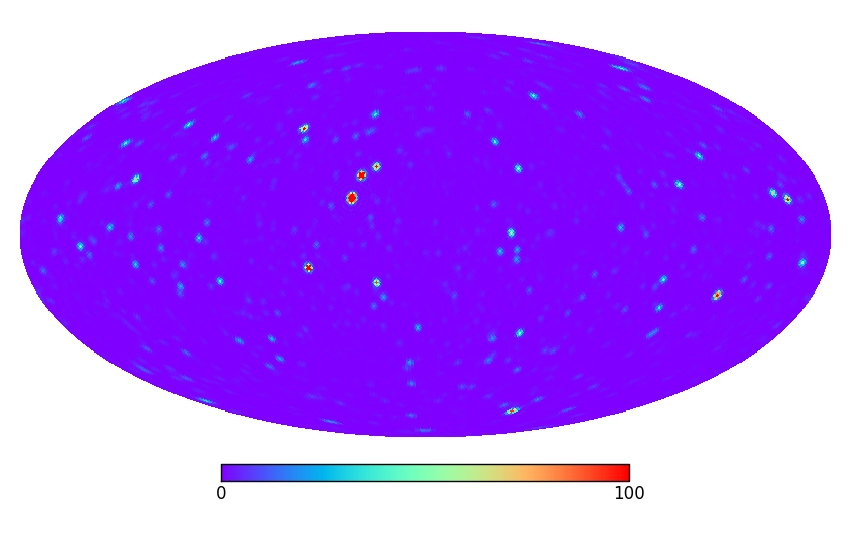} 
\includegraphics[clip=true,angle=0,width=0.48\textwidth]{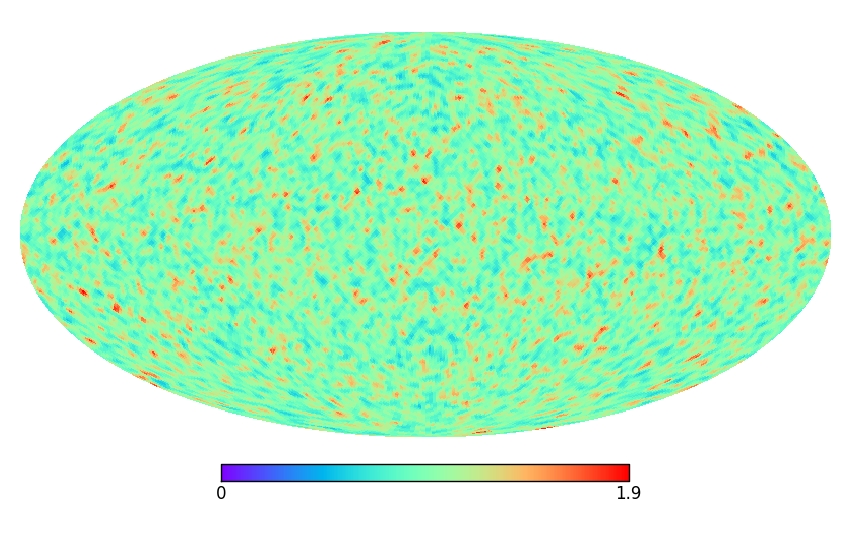} 
\includegraphics[clip=true,angle=0,width=0.48\textwidth]{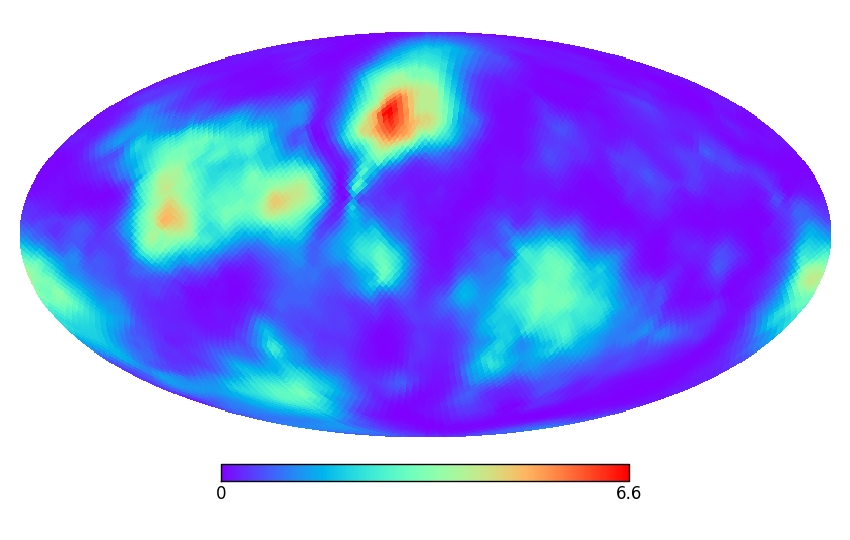} 
\includegraphics[clip=true,angle=0,width=0.48\textwidth]{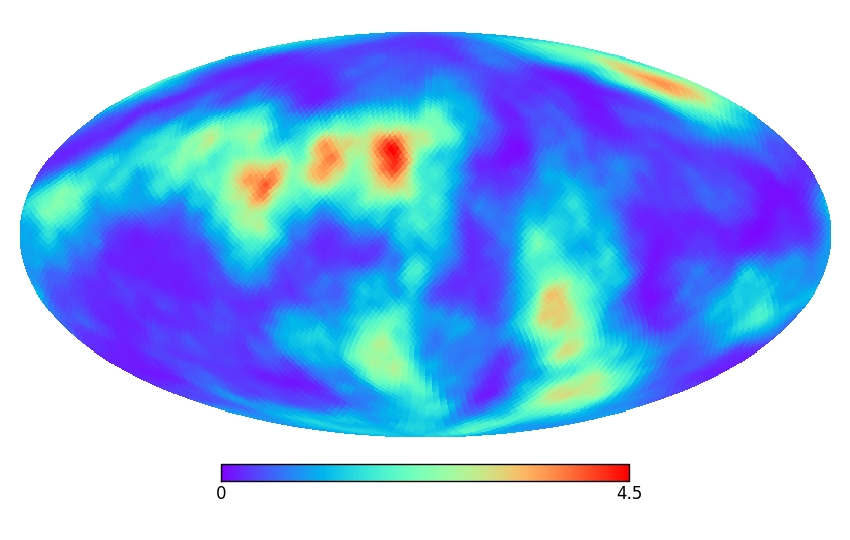} 
\caption{\label{fig:skies} The upper two maps show the distribution of the gravitational wave power across the sky, scaled by the all-sky average for two signal model. The map on the upper left is for a
simulated black hole population, smoothed with a two degree Gaussian blur and clipped at a contrast of 100 to enhance weaker features. The
peak intensity in the un-clipped map exceeds 800. The map on the upper right is for a statistically isotropic signal with the same power spectrum as the black hole simulation. The map has been smoothed with
a two degree Gaussian blur. Clearly, the power distribution from a realistic black hole population is far from isotropic. The lower two panels show the detected power in the Earth-term for pulsars at different sky locations, in other
words, the raw signals convolved with the antenna patterns, summed and squared. Despite the large differences in the underlying power distribution, the response to the anisotropic BH background (lower left) is
qualitatively identical to the response to the statistically isotropic signal (lower right). }
\end{figure*}

Electromagnetic observations of massive galaxies and galaxy mergers across cosmic history, combined with population synthesis models, suggest that the dominant source of gravitational
waves in the pulsar timing band ($10^{-9}\; {\rm Hz} \rightarrow 10^{-6}\; {\rm Hz}$) will be slowly evolving supermassive black hole binaries with masses in the range
$10^8 M_\cdot \rightarrow 10^9 M_\cdot$~\cite{2003ApJ...590..691W,2003ApJ...583..616J,1995ApJ...446..543R}. It was initially assumed that the superposition of the signals from many thousands of such systems would produce a background
that is effectively stochastic and statistically isotropic, and thus amenable to detection using the cross-correlation technique developed by Hellings and Downs~\cite{HellingsDowns}. Recent studies of the signals produced by simulated black hole populations, though, have shown that relatively nearby and massive outliers play
an important role, and can lead to significant departures from stochasticity and isotropy~\cite{2014MNRAS.439.3986R,2012ApJ...761...84R}. 

To illustrate the importance of outliers in these populations, we simulate the gravitational wave signals
from a population model provided by A. Sesana that assumes quasi-circular, gravitational wave driven orbital evolution. The gravitational waves from each binary are co-added and used to compute
the signal power as a function of sky position, $h^2_{\rm ss}(\theta,\phi) = h_+^2(\theta,\phi)+h_\times^2(\theta,\phi)$, summed over frequency. 
Figure 1 shows the distribution of the gravitational wave power across the sky for one realization of the black hole population, as well as for a statistically isotropic stochastic background with the same
average power spectrum. The difference is striking. The intensity variations for the black hole population are over one hundred times larger than for the isotropic model. The lower two panels of Figure 1
show the pulsar response to the signals as a function of pulsar sky location. (Only the Earth-term contribution is shown here. Including the pulsar terms simply adds ``noise'' to the maps.) Note that
the pulsar response has a similar angular power distribution for both the isotropic and black hole skies, even though the underlying signals are vastly different.  The explanation can be found in Figure 2, which shows
the detected power in pulsars at different sky locations for a single black hole binary. The broad antenna response of the pulsars effectively blurs the underlying power distribution. Note, however,
that this does not imply that pulsar timing arrays are unable to resolve small scale features - the information to reconstruct the spatial distribution resides in the cross-spectra, and this information
can be used to accurately map the background~\cite{2013PhRvD..88f2005M,2013PhRvD..88h4001T,2015PhRvL.115d1101T,2014arXiv1406.4511C}.

\begin{figure}[htp]
\includegraphics[clip=true,angle=0,width=0.48\textwidth]{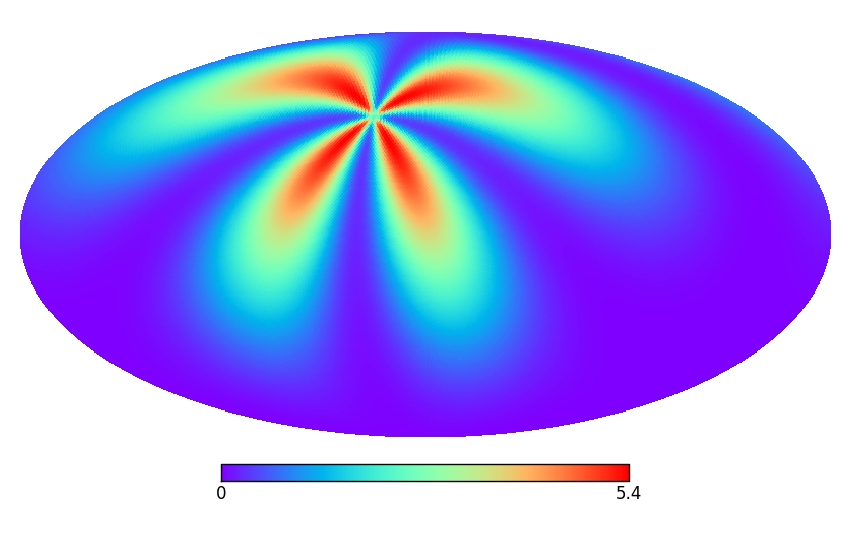} 
\caption{\label{fig:single} The detected power for pulsars at different sky locations for a single BH located at the center of the ``petals''. The orientation of the petals rotates depending on the polarization
of the signal. The broad point spread function for PTAs explains why the response to isotropic and anisotropic signals is so similar.}
\end{figure}

Our procedure for testing the tensor correlation analysis on realistic black hole populations and pulsar timing arrays is to compare the evidence for detections between simulated black hole populations and
statistically isotropic signals with the same average power level. To do this we first simulate the response to a particular realization of the black hole population model for an array of pulsars, and from this
compute the average power spectrum (averaged over the pulsars), $S_h(f)$.  We then simulate a statistically isotropic, unpolarized Gaussian stochastic background with this same power spectrum.
For reference, we also consider the standard
power law spectrum for gravitationally wave drive, quasi-circular binaries which have characteristic strain $h_c(f) = A (f / f_y)^{-2/3}$ and $S_h(f) = A^2 (f / f_y)^{-4/3} /(12 \pi^2 f^3)$ where
$f_y = 1/{\rm year}$. Figure 3 shows examples of the average power spectrum for the three models for a 20 pulsar array. The power law model is generated with $A=10^{-15}$. The slight differences in the black hole
spectrum and the equivalent isotropic spectrum are due to the fact that this is a single realization of the stochastic power spectrum, averaged over the array.

\begin{figure}[htp]
\includegraphics[clip=true,angle=0,width=0.48\textwidth]{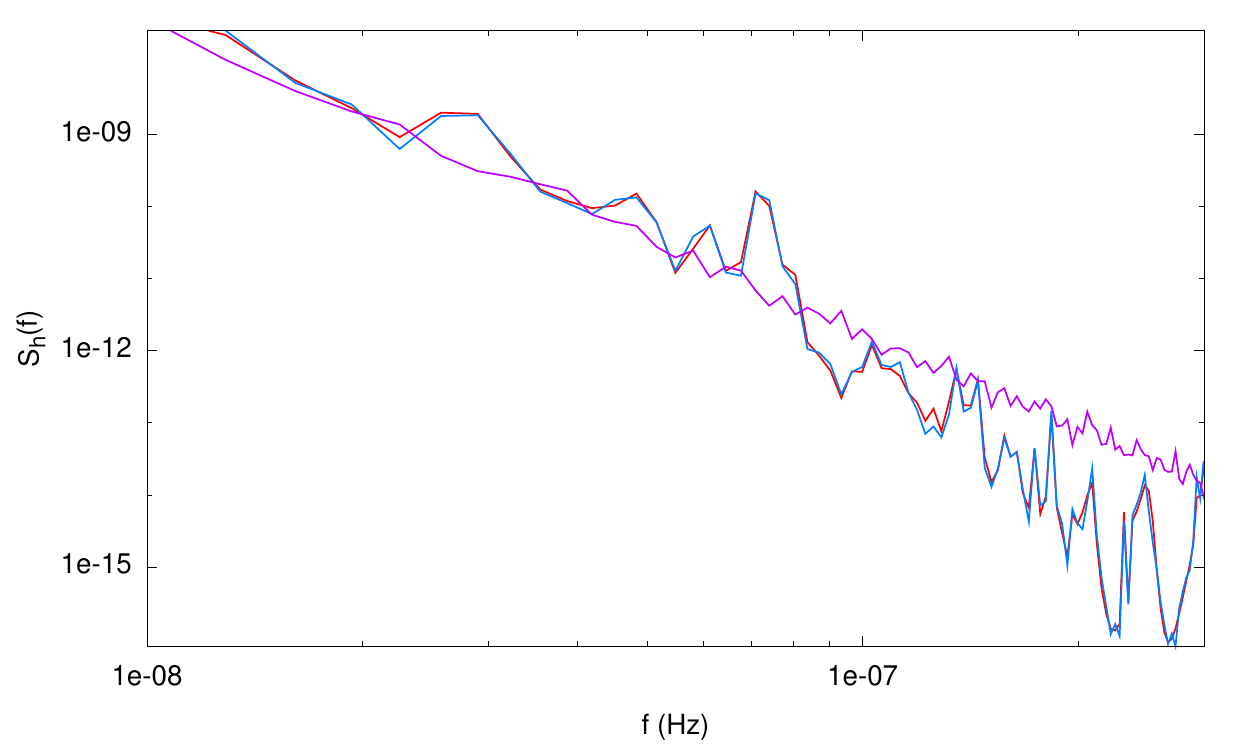} 
\caption{\label{fig:power}   The average power spectrum in a  20 pulsar array for a black hole population (in red), a stochastic, isotropic model with the same average power level as the black hole population (in blue),
and a  power law, stochastic, isotropic model with amplitude $A=10^{-15}$.}
\end{figure}

The reference astrophysical model we are using - based on a quasi-circular, gravitational wave driven merger model - likely {\em overestimates} the degree of isotropy we can expect in reality. Environmental effects, such as stellar scattering and gas driven mergers, along with orbital eccentricity will act to reduce the number of binary systems in each frequency bin. As a crude proxy for these effects, we also produce simulated 
backgrounds that randomly down-sample the full black hole population by factors of 10 and 100, and, as an extreme case, produce backgrounds that include only the 10 brightest sources from the full simulation (in contrast, the
full population model includes over 22,000 sources in the observation band). In addition to considering different numbers of sources in the signal, we also investigate the impact of including different numbers of pulsars in
the array. We investigate simulated arrays with 5 equally sensitive pulsars, 20 equally sensitive pulsars, and the 36 pulsars of varying sensitivity taken from the first International Pulsar Timing Array (IPTA) mock
data challenge. The simulated data sets include white noise at a range of levels, but no simulated red noise or residuals from the quadratic spin-down model. The effects of red noise and timing errors are, however,
included in the analysis.

\section{Analysis}
\label{sec:Analysis}

We apply Bayesian inference and model selection to analyze the simulated data sets. The analysis procedure is very similar to that described in Ref.~\cite{Sampson:2015ada}.  
The likelihood of observing data $d$ for a given model set of model parameters $\vec{\lambda}$ is 
\begin{equation}
p(d|\vec{\lambda}) = \frac{\exp \left(-\frac{1}{2} \sum_{ab} \sum_{ij} r_{ai} C^{-1}_{(ai)(bj)} r_{bj}  \right)}{\sqrt{(2\pi)^M \det C}} ,
\end{equation}
where C is the covariance matrix, which depends on both the noise in the individual pulsars and on the GW background, and $r= d-t$ denotes the timing residuals after the subtraction of the timing model $t$ from the data $d$. The indices $a$ and $b$ label individual pulsars, and run from $1$ to the number of pulsars, $N_p$. The indices $i$ and $j$ label the data samples, i.e. individual frequency bins. Since our simulated data is stationary, the
correlation matrix is diagonal in $i,j$ and $C_{(ai)(bj)} \rightarrow C_{ab}(f_i)\delta_{ij}$. The simulated data set consists of $N=512$ samples per pulsar, evenly spaced in time at weekly intervals, giving a total data set of size
$M=N \, N_p$ spanning just under  $T=10$ years. The analysis is carried out in the Fourier domain, where the quadratic timing model for pulsar $a$ has the form
\begin{equation}
t_a(f_k) =  \frac{\alpha_a }{f_k^2} + \frac{i \beta_a}{f_k},
\end{equation}
where $f_k = k/T$ for integer $k$. 
The covariance matrix is given by
\begin{equation}\label{cmatrix}
C_{ab} (f) = S_h(f)H_{ab} + \delta_{ab} \left\{S_{n_a} + S_{r_a} (f/f_y)^{r_a}\right\},
\end{equation}
where $S_h(f)$ is the PSD of the GW background, $S_{n_a}$ is the PSD of the white noise, $S_{r_a}$ is the amplitude of the PSD of the red noise, and $r_a$ is the spectral slope of the red noise (which should not be confused with the $r_a$ from Eq. (2), which represents the residuals in pulsar $a$!).
In the sky-scramble analysis the tensor correlation matrix, $H_{ab}$, is replaced by a scrambled version that is derived by randomly choosing false sky locations for each pulsar.
We consider two models for $S_h(f)$, a simple power law $S_h(f) = S_g (f/f_y)^{\gamma}$, and a more general bin-by-bin model $S_h(f_k) = P_k$, where $P_k$ is the power spectral density in the $k^{\rm th}$
frequency bin. There are $N/2$ of these $P_k$ parameters, given an observation time of $T$ and a Nyquist frequency of $0.5/dt$, so $N_{\rm{bin}} = 0.5*T/dt =N/2.$The bin-by-bin model has many more parameters than the power-law model, and is far less effective at detecting a power-law signal, but it has added flexibility, and can better
capture the non--power-law spectra that arise for the very sparse black hole population models. The full parameter vector $\vec{\lambda}$ for the timing plus noise model has the $5N_p$ parameters
$\vec{\lambda}\rightarrow \{\alpha_a, \beta_a, S_{n_a}, S_{r_a}, r_a \}$, while the power-law gravitational wave model has the $2+5N_p$ parameters
$\vec{\lambda}\rightarrow \{ S_g, \gamma, \alpha_a, \beta_a, S_{n_a}, S_{r_a}, r_a \}$, and the bin-by-bin gravitational wave model has the $N/2+5N_p$
parameters $\vec{\lambda}\rightarrow \{P_k, \alpha_a, \beta_a, S_{n_a}, S_{r_a}, r_a \}$. The priors $p(\vec{\lambda})$ on the power spectral density parameters $\{ S_g, P_k, S_{n_a}, S_{r_a}\}$ are taken to be uniform
in the logarithm across the range $[10^{-35} {\rm Hz}^{-1}, 10^{-4} {\rm Hz}^{-1}]$. The priors on the spectral slope parameters $\{\gamma, r_a\}$ are taken to be uniform in the range $[-2,-6]$, and the
priors on the timing model parameters $\{\alpha_a T^2, \beta_a T\}$ are taken to be uniform in the range $[-0.8,0.8]$.

In the analyses that follow we are less interested in the posterior distributions for the model parameters, $p(\vec{\lambda} | d)$, than we are in the model evidence
$p(d) = \int p(d|\vec{\lambda}) p(\vec{\lambda}) d\vec{\lambda}$ for the various models. In particular, we compute the Bayes factor for a detection as the evidence ratio
between the signal model and the noise model. The evidence is computed using the thermodynamic integration technique~\cite{2004AIPC..707...59G}, which returns the evidence as a natural by-product
of the parallel tempered Markov Chain Monte Carlo scheme~\cite{1986PhRvL..57.2607S} that we use to map the posterior distributions. The implementation of the MCMC algorithm is as described in
Ref.~\cite{Sampson:2015ada}, with two additional features: an additional move that proposes to transfer power between the signal and red noise models, and an adaptive scheme for the temperature ladder.

The adaptive scheme is as follows: we begin with $50$ chains equally spaced between in log temperature between $1$ and $10^6$, and perform an MCMC run with this spacing while keeping track of the acceptance rate for parallel tempering moves between all adjacent pairs of chains. If one of these acceptance rates falls below $1\%$, we stop the run and insert 3 chains with temperatures evenly spaced between the two chains that lost contact. We continue this process until the MCMC runs for $10^6$ iterations without adding any chains. We then use this final temperature ladder (usually including $\sim90-100$ chains) to perform the evidence calculation, using an MCMC run of $1.5$ million iterations.

As indicated, the new move we have implemented shifts power between the GW signal and the independent red noise in each pulsar. To do this, we calculate the average level of red noise in all of the pulsars, $\bar{S}_r = \sum_i S_r ^i /NP$, and propose that the GW amplitude takes this value by proposing from a Gaussian centered at this value with a width of $\sigma_1 = 0.5$. We simultaneously propose that the red noise level in each pulsar be drawn from a Gaussian centered at the current red noise level, with a width of $\sigma_2 = \sigma_1/\sqrt{NP}$. We find that this proposal greatly aids the mixing between models. A nearly identical proposal can be used to move power between the red noise in the individual pulsars and the common red noise level, if such a term is present in the model.

\section{Detecting anisotropic backgrounds}

Several methods have been proposed for detecting and mapping anisotropic gravitational wave backgrounds with pulsar timing
arrays~\cite{Cornish:2013aba, 2013PhRvD..88f2005M,Taylor:2013esa,Gair:2014rwa, Cornish:2014rva}, but it remains to be seen if these methods are
more effective at making a first detection than the standard tensor correlation analysis. In the case of the cosmic microwave background radiation, the uniform glow was detected long before the first
anisotropies were seen, but in that case the anisotropies are tiny compared to the overall power. The much larger anisotropy of the nanoHertz gravitational wave sky may mean that a model that allows for anisotropy will
improve the prospects for detection. But it is not obvious that this will be true, because while the data might be better fit by an anisotropic model, such models are necessarily more complicated than the isotropic model, and this added complexity
comes at a price. We defer the comparison of the efficacy of isotropic and anisotropic models for future study, and instead consider the simpler question of how effective the standard isotropic
analysis is when applied to realistic anisotropic signals using realistic numbers of pulsars in the array. We accomplish this by comparing the detectability of anisotropic signals from a population of
black holes to the detectability of a statistically isotropic background with an identical power spectrum. 

We consider two measures of detectability, the first being the Bayes factor, or evidence ratio, between the signal and noise models. Recall that the noise model includes individual red and white noise
components for each pulsar, which are uncorrelated between pulsars, and the signal model includes a common stochastic component with a red power spectrum, with the characteristic Hellings-Downs correlation pattern between
pulsars. Unfortunately, as we show in section \S\ref{skyscram}, this signal-to-noise model Bayes factor will imply the detection of signals that do not have the correct tensor correlation pattern. (Though with
lower significance than similarly bright signals that do.) Because the tensor correlation pattern is key to any claim of a gravitational wave detection with pulsar timing, we go on to consider a second
measure of detectability - this measure compares the evidence for the tensor correlation model to a model with a diagonal correlation matrix. This diagonal model corresponds to a common level of red noise present in all pulsars.

\begin{figure}[htp]
\includegraphics[clip=true,angle=0,width=0.48\textwidth]{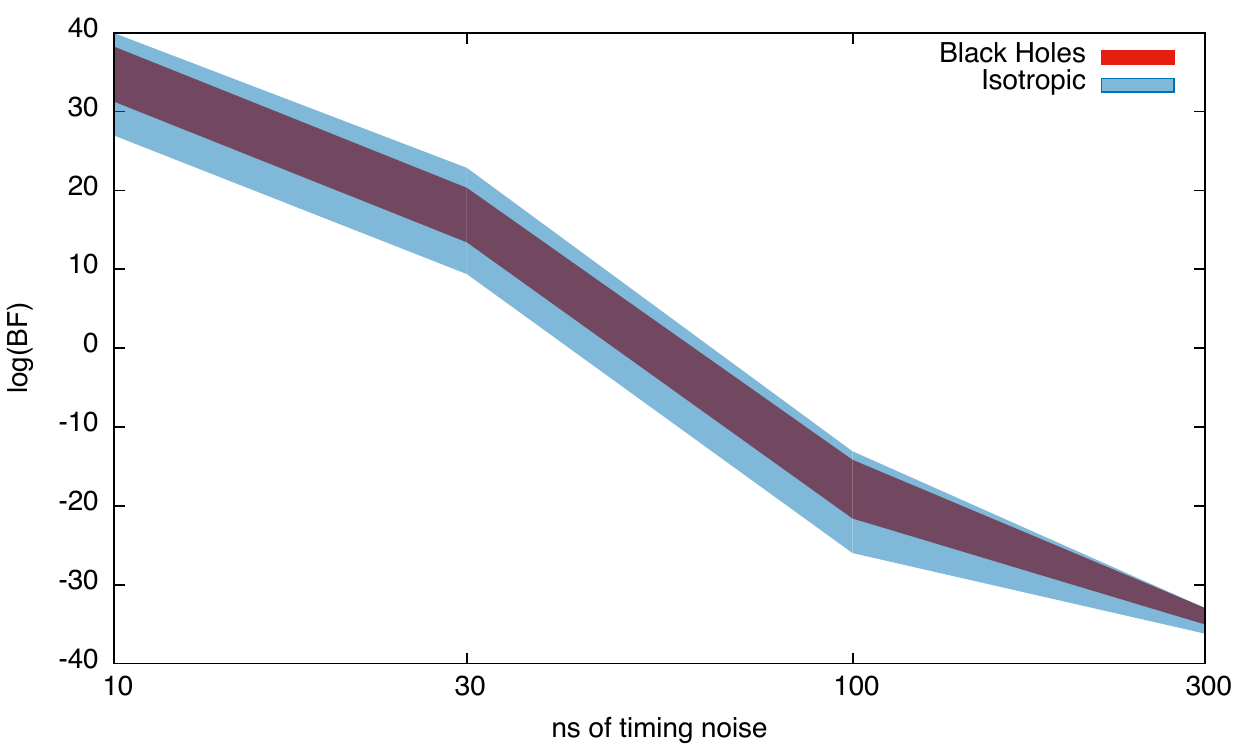} 
\caption{\label{fig:20full}  The detectability of the full simulated black hole population (red) and a statistically isotropic stochastic signal with the same power spectrum (blue) as a function of the
white timing noise level in a simulated pulsar timing array with 20 equally sensitive pulsars, as measured by the signal-model to noise-model Bayes factor.  The spread in the Bayes factors is
computed by consider 10 realizations of each signal type. The shaded bands covers a
one standard deviation spread about the mean.}
\end{figure}

Even within one class of signals there is considerable variation in detectability from realization to realization. To account for this we consider multiple realizations for each signal type and aggregate the
results. Figure 4 compares the detectability of the signal from a full black hole population to that of a statistically isotropic stochastic signal with the same power spectrum, as a function of the
white timing noise level in a simulated pulsar timing array with 20 equally sensitive pulsars. (In this figure we define detectability in terms of the evidence ratio between the signal and noise models).
Remarkably, we see that there is no discernible difference between the detectability of the two types of signal,
even though the black hole signal is far from isotropic and the array is comprised of relatively few pulsars.

\begin{figure}[htp]
\includegraphics[clip=true,angle=0,width=0.48\textwidth]{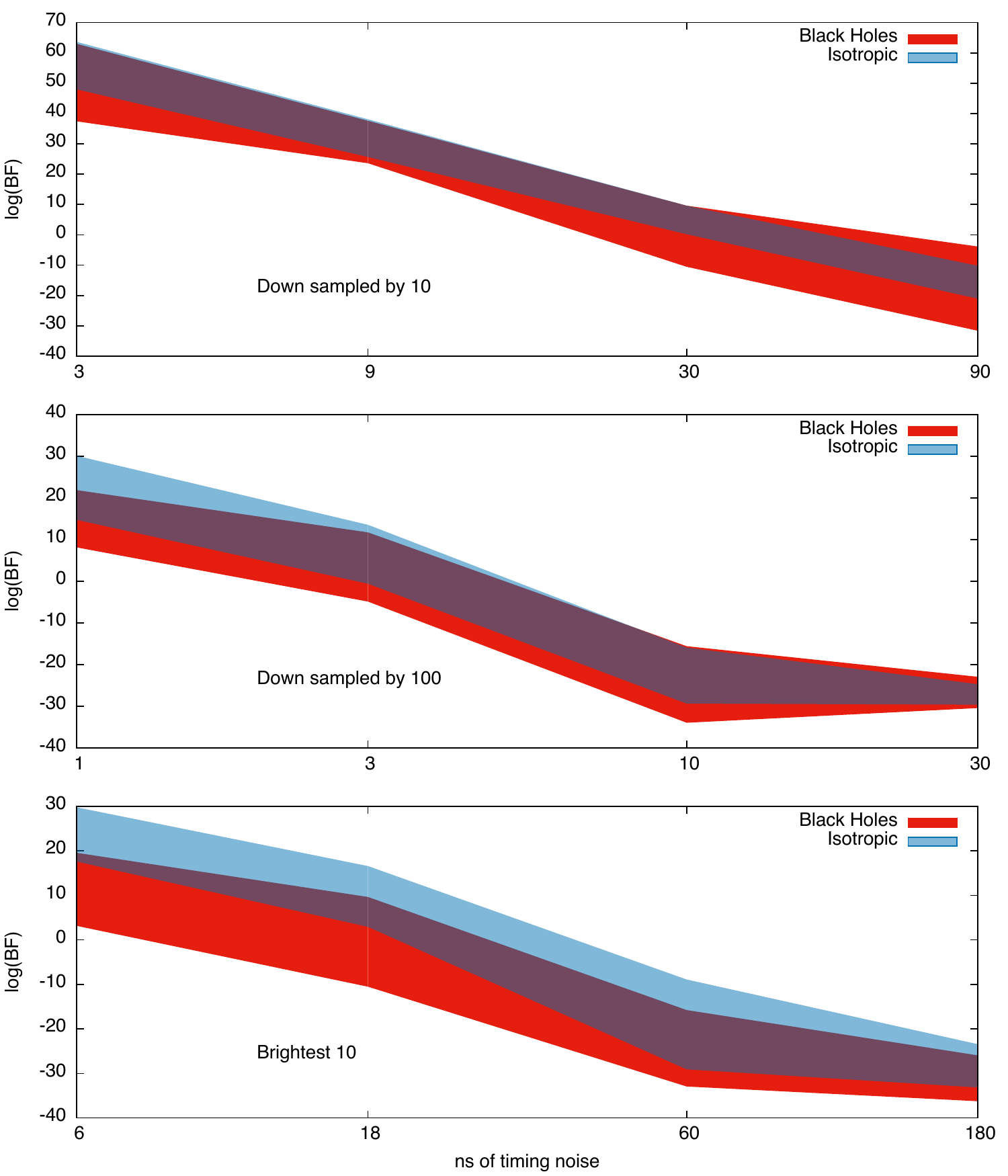} 
\caption{\label{fig:20down}  The detectability of down sampled black hole populations (red) and a statistically isotropic stochastic signal with the same power spectrum (blue) as a function of the
white timing noise level in a simulated pulsar timing array with 20 equally sensitive pulsars, as measured by the signal-model to noise-model Bayes factor.  The upper panel is for the full population
down sampled by a factor of ten, the middle panel is down sampled
by one-hundred, and the lower panel is for the ten brightest binaries. The spread in the Bayes factors is computed by consider 10 realizations of each signal type. The shaded bands covers a
one standard deviation spread about the mean.}
\end{figure}

Figure 5 shows similar results as Figure 4, but for smaller black hole populations (i.e., more anisotropic skies). Once again the anisotropic signals are almost as detectable as their isotropic equivalents. Differences only become apparent
for exceedingly sparse black hole populations with only a handful of sources. One may be concerned that the simulations that use only the ten brightest black holes to generate the background may not
be well described by a power law. Examining the power spectra for these cases, though, shows that they can indeed be fit by power laws, although typically with slopes that are steeper than for the full
population. We additionally investigate how the size of the array affects the results by considering a smaller array made up of 5 equally sensitive pulsars. The upper panel in Figure 6 shows the
detectability of the full black hole population and its isotropic equivalent, while the lower panel in Figure 6 shows a more extreme case, with just 10 black hole binaries and 5 pulsars in the array.
In this case the correlation analysis does perform worse on the black hole population than on the isotropic equivalent, but the difference is still within the uncertainty from realization to realization.

 \begin{figure}[htp]
\includegraphics[clip=true,angle=0,width=0.48\textwidth]{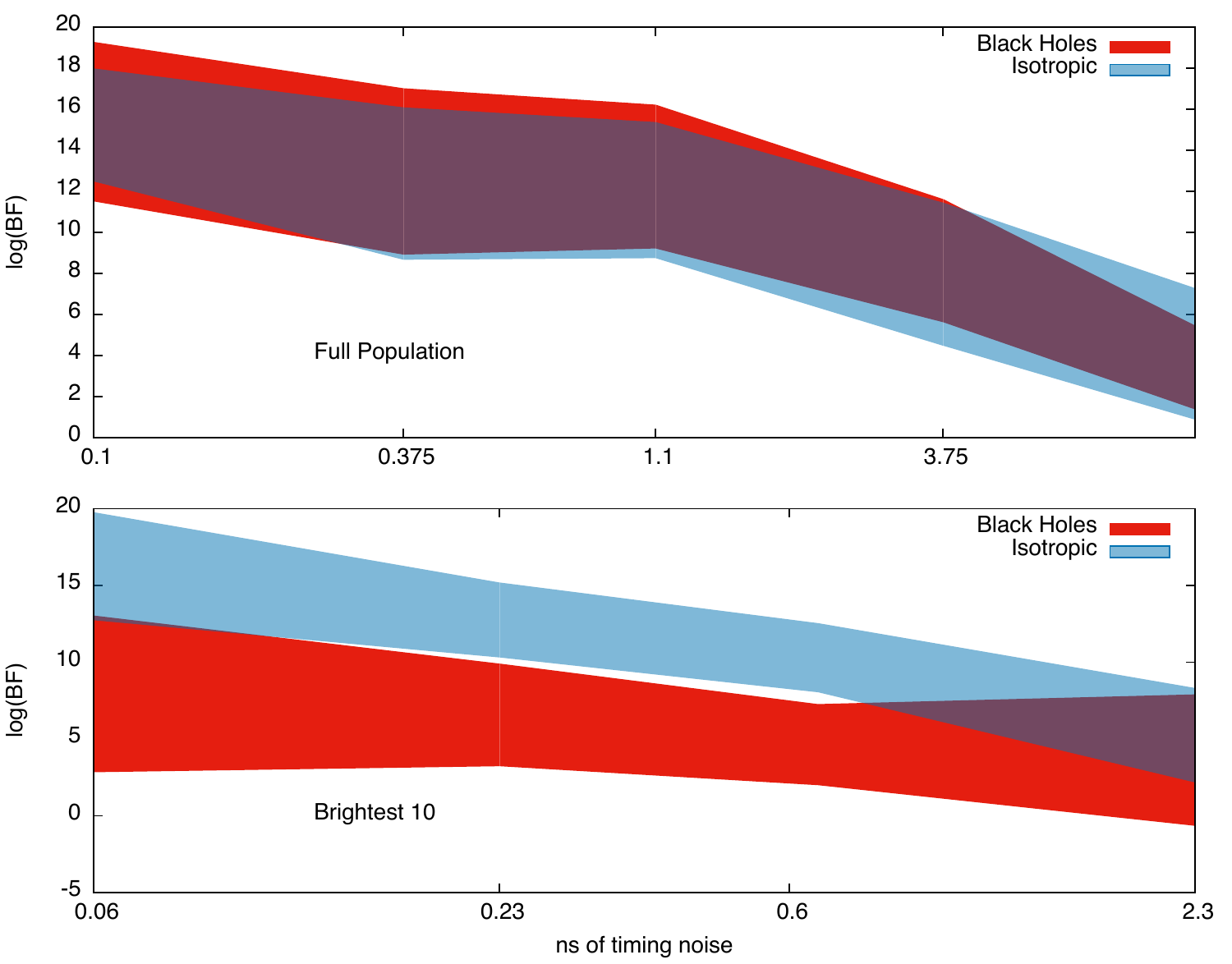} 
\caption{\label{fig:20down}  The detectability of signals from a population of black holes (red) and a statistically isotropic stochastic signal with the same power spectrum (blue) for
a small pulsar timing array made up of 5 equally sensitive pulsars, as measured by the signal-model to noise-model Bayes factor. The upper panel is for the full black hole population,
while the lower panel is for the brightest 10 black holes.
The spread in the Bayes factors is computed by consider 10 realizations of each signal type. The shaded bands covers a one standard deviation spread about the mean.}
\end{figure}

 Finally, we explore the effect of anisotropy on the detection of a stochastic GW background using the pulsars from the IPTA. Instead of injecting a particular noise level into all of the pulsars in the array, we scale the actual noise level for each of the pulsars by the same factor. One might expect that this array would behave much like the 20 pulsar array explored previously in this section. It turns out, though, that there are a few pulsars in the array that are much better-timed than the others, and so dominate detection. Figure 7 shows the results from this study, presented in the same format as previous results in this section. The results are much more like the 5 pulsar array than the 20 pulsar case.

\begin{figure}[htp]
\includegraphics[clip=true,angle=0,width=0.48\textwidth]{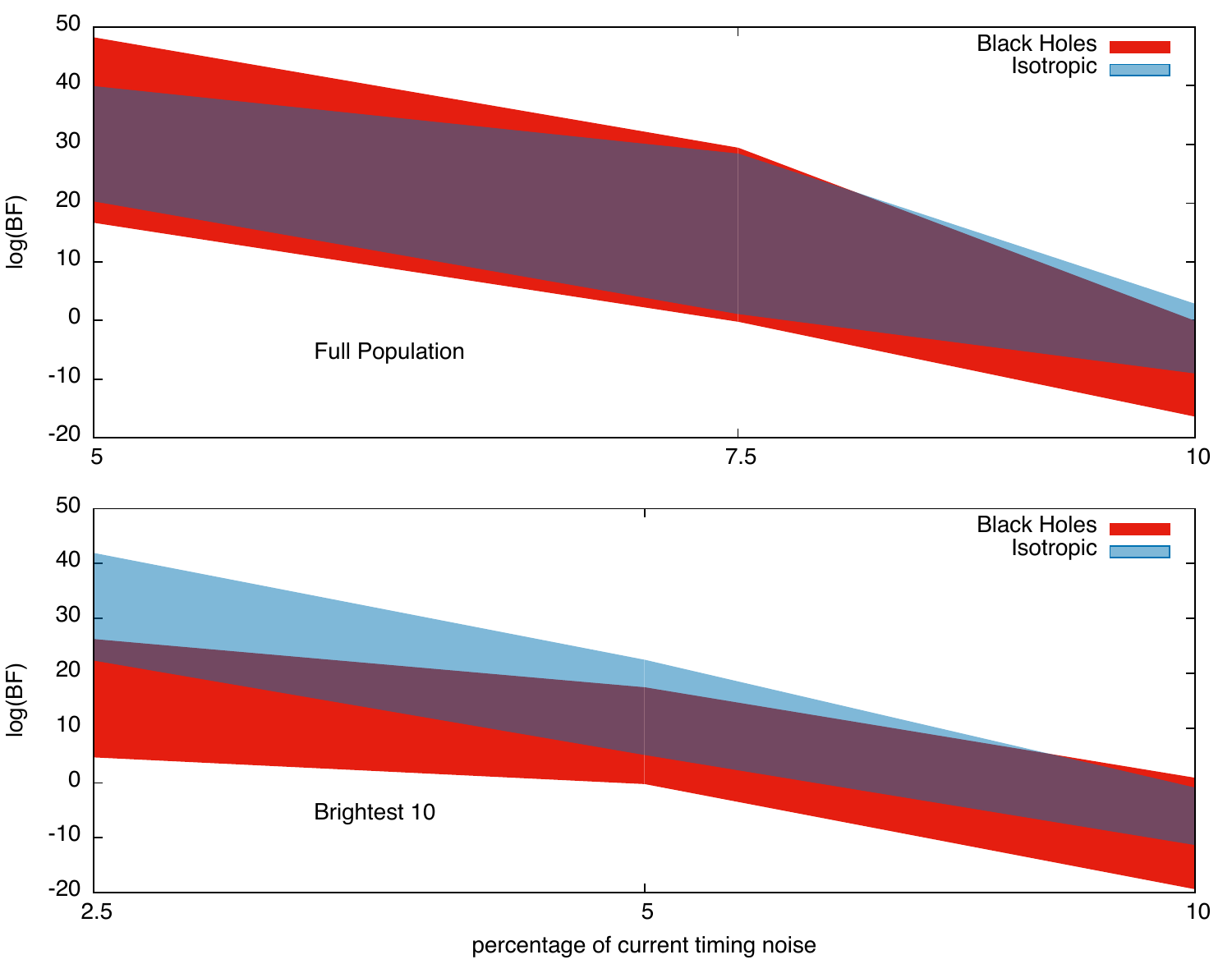} 
\caption{\label{fig:20down}  The detectability of signals from a population of black holes (red) and a statistically isotropic stochastic signal with the same power spectrum (blue) for
the IPTA pulsar array, again for the full spectrum in the upper panel and the brightest 10 black holes in the lower panel. The horizontal axis shows the percentage of the actual timing noise in each pulsar that is included in the injected data. The results are more similar to the 5 pulsar array than the 20 pulsar case, despite the fact that there are 36 pulsars in the IPTA. This is because a few of the pulsars are much better timed than the others.}
\end{figure}

While our analysis indicates that the standard correlation analysis is almost as effective at detecting anisotropic signals as it is at detecting the isotropic signals it was designed for, it is unclear if the signal model is
picking up the full tensor correlation pattern (\ref{tensor}), or merely a common red component in the timing residuals - i.e. just the diagonal terms in (\ref{tensor}). While the gravitational wave signal model includes a diagonal component, there could conceivably be correlated power on the diagonal due to some common red noise process. For example, there may be some physical process that is shared by all neutron stars that produces a
characteristic spectrum of red timing noise~\cite{Shannon:2010bv}. To be sure that it is a gravitational wave signal that has been detected we need clear evidence that the off-diagonal, cross-correlation terms
follow the Hellings-Downs curve. One approach is to try and infer the correlation pattern directly from the data, using techniques such as a cubic spline fit to the correlation pattern~\cite{Taylor:2012wv}, or by directly
inferring the correlation of each pulsar pair~\cite{Lentati:2012xb} and comparing this to the Hellings-Downs curve. Another approach is to apply Bayesian model selection between a model with the full tensor correlation
curve $H_{ab}$ and a model with a common red noise term, given by the diagonal correlation model $H'_{ab}= \delta_{ab}$. We opt to follow the later approach, which was first described in Ref.~\cite{Ellis:2013nrb}.
The common red noise term can either be considered in addition to the gravitational wave model as an extra term in the noise model, or by comparing ``signal'' models with correlation matrices given by
$H_{ab}$ and $H'_{ab}$. We settle on the latter approach, as including an extra common red noise model made it very difficult to compute reliable estimates for the evidence. This is because the large correlations
between the common red noise model, the per-pulsar red noise model, and the signal model impeded mixing of the Markov chains, and led to a series of steep transitions in the thermodynamic integration integrand.
Moreover, even with reliable evidence estimates produced by using vast numbers of steps in the temperature ladder, the results are highly dependent on the choice of the priors on the various parameters
in each model, which is always an issue when comparing models of different dimension. Comparing the full and diagonal correlation models is much easier, and the choice of priors has much less of an effect, as
the two models share the same parameters, effectively canceling the prior dependence in the evidence ratios. In the language of Ref.~\cite{Ellis:2013nrb}, we are comparing the evidence of models $M_{\rm gw}$
and $M_{\rm corr}$, whereas our earlier results compared the gravitational wave model $M_{\rm gw}$ to the noise model $M_{\rm null}$. We demand that the evidence for $M_{\rm gw}$ exceeds the evidence for
both $M_{\rm null}$ and $M_{\rm corr}$ to claim a detection. Figure 8 shows the Bayes factors between the full tensor correlation model $M_{\rm gw}$ and the noise model $M_{\rm null}$, and the Bayes factors
between the diagonal correlation model $M_{\rm corr}$ and the noise model $M_{\rm null}$ for a simulated pulsar timing array with 20 equally sensitive pulsars. Each panel shows the results for an increasingly
anisotropic signal from a population of black holes, with the spread in Bayes factors computed by considering multiple realizations of each population to account for cosmic variance.

\begin{figure}[htp]
\includegraphics[clip=true,angle=0,width=0.48\textwidth]{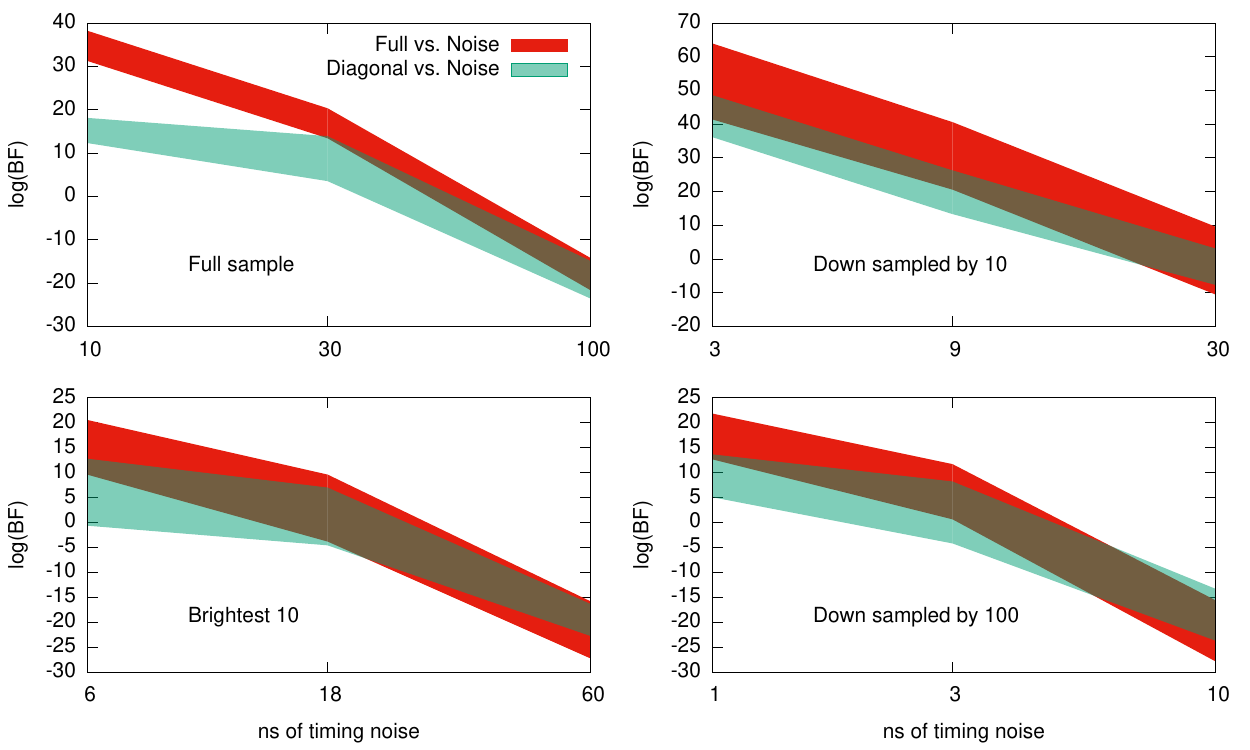} 
\caption{\label{fig:fulldiag}  The detectability of simulated black hole populations as a function
of the noise level for a simulated pulsar timing array with 20 equally sensitive pulsars. The red bands
show the spread in log Bayes factors for the full tensor correlation model computed from multiple realizations of a black hole population model. The green bands show the spread in log Bayes factors
for the diagonal correlation model applied to the same set of simulated signals. The gravitational wave signal model is favored when the log Bayes factor for the tensor correlation model versus the noise
model is positive {\em and} that it exceeds the log Bayes factor for the diagonal correlation model versus the noise model. From top to bottom the simulations are for the full black hole background,
the populations down-sampled by ten then one hundred, and finally for just the ten brightest systems.}
\end{figure}

Figure 9 shows the log Bayes factors between the full tensor correlation model $M_{\rm gw}$ and the diagonal common noise model $M_{\rm corr}$, for both anisotropic black hole populations and their
isotropic equivalents, with the spread showing the cosmic variance derived by considering multiple realizations. Values greater than zero indicate that
the Hellings-Downs correlation curve has been detected in the data. As expected, the evidence ratio for the full and diagonal models tends to unity as the noise level increases, because these models have
the same prior volume. We see that the results for isotropic and anisotropic signals are essentially indistinguishable, which implies that not only is a common red noise component being detected in
both cases, but so is the tensor correlation pattern.

\begin{figure}[htp]
\includegraphics[clip=true,angle=0,width=0.48\textwidth]{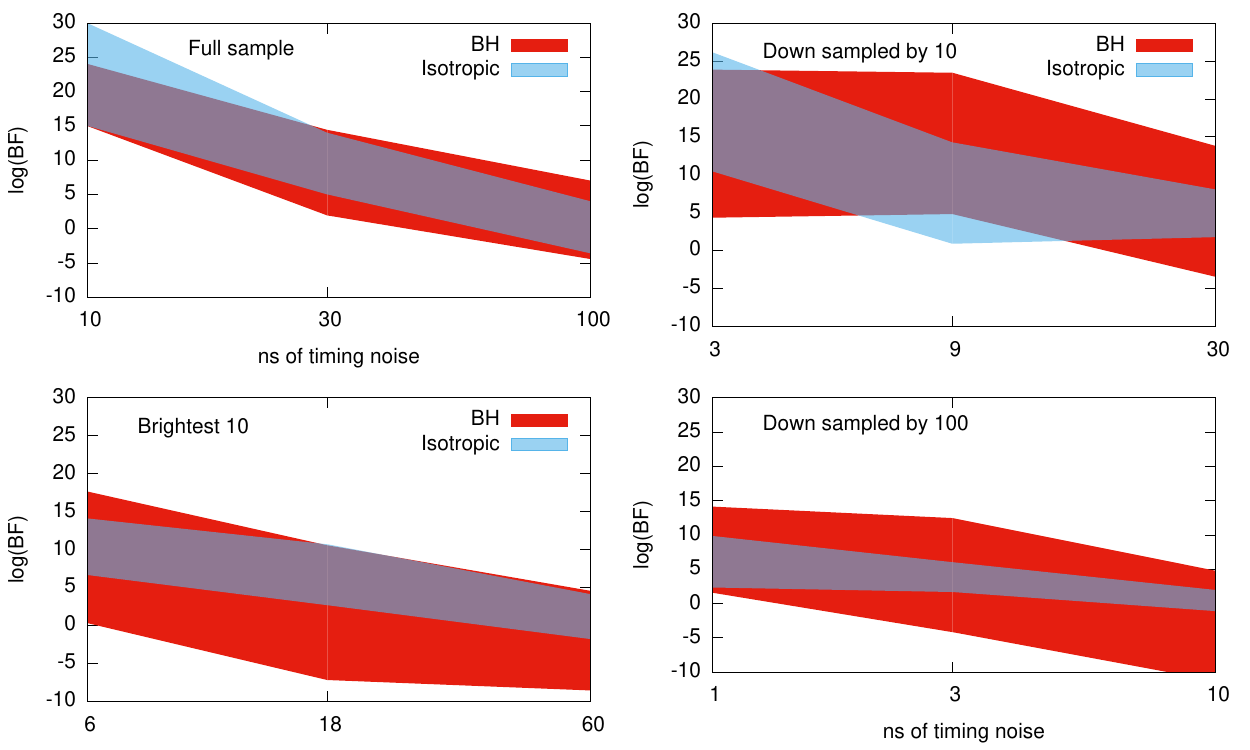}
\caption{\label{fig:fulldiag2} The log Bayes factors between
the full and diagonal correlation model. Values greater than zero indicate that the Hellings-Downs correlation pattern has been detected. The red bands are for
simulated black hole populations, and the blue bands are for the isotropic equivalents. }
\end{figure}

\section{Sky Scrambles}\label{skyscram}

We have found the tensor correlation pattern to be a remarkably robust signature for detecting gravitational waves with pulsar timing arrays. As the existing pulsar timing arrays continue to
collect data at improved sensitivity, and as more pulsars are added to the arrays, we should start to see the first hints of correlated gravitational wave power in the timing residuals~\cite{Siemens:2013zla}.
The evidence for a signal will then grow steadily with time, until eventually the evidence becomes overwhelming. However, the evidence we compute is between {\em our model for the signal} and
{\em our model for the noise}, and deficiencies in either of these models could lead to false positives or false negatives. Assuming that general relativity provides a faithful description of gravity in
the regime probed by pulsar timing, our model for the signals and how they perturb the timing residuals should be reliable, but the many potential sources of noise are less well understood. Ideally,
we would like to be able to study the noise properties in data that is free of gravitational waves, but there is no way to shield our detector from gravitational wave signals. 

The same challenge occurs in the analysis of data from the ground-based LIGO/Virgo interferometers~\cite{TheLIGOScientific:2014jea, Accadia:2015pda}, where studies have shown the noise to be
both non-stationary and non-Gaussian~\cite{Aasi:2012wd, Cornish:2014kda, Littenberg:2014oda}, with frequent
loud transient features, or glitches~\cite{Blackburn:2008ah}. While it is not possible to remove gravitational wave signals from the data, it is possible to destroy signal
coherence across the detector network by introducing artificial time
delays between the detectors during the analysis~\cite{Was:2009vh}. Because the noise in each detector is assumed to be uncorrelated to begin with, the time slides preserve the statistical properties of the noise. The groups that
analyze the data from ground based detectors approach the detection problem in a frequentist framework, using some detection statistic to identify candidate events. The distribution of triggers from the time slides
are used to compute false alarm rates, and any triggers in the zero-lag data that exceed a pre-ordained false alarm rate (such as one per millennium) are deemed detection candidates. To establish a false alarm
threshold of one per millennium for a year long data set requires thousands of independent time slides. This can be achieved for the ground-based detectors, as the correlation length of typical signals in the ground based interferometer band are generally less that a second. 

\begin{figure}[htp]
\includegraphics[clip=true,angle=0,width=0.48\textwidth]{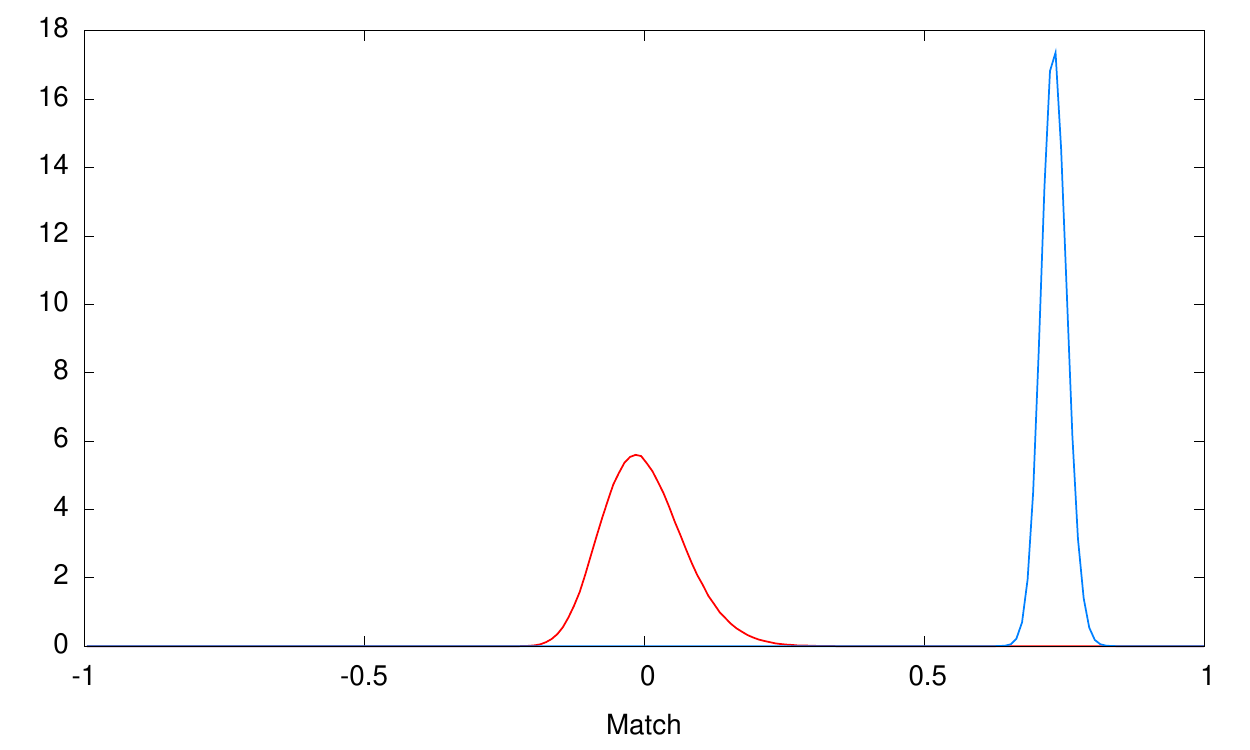} 
\caption{\label{fig:scram}   Histograms of the full match $M$ (in blue), and the off-diagonal match $\overline{M}$ for randomly drawn sky-scrambles of an equal-sensitivity 20 pulsar timing array}
\end{figure}

In the case of pulsar timing, thoough, the correlation length of the gravitational wave signals is expected to be comparable to the duration of the available data sets, making
it impossible to generate independent time-slides. Instead, we can break the \emph{spatial} correlations by artificially scrambling the pulsar positions used in the gravitational wave analyses (the true positions still
have to be used in the timing model). We can define the match between two sets of correlation matrices $H_{ab}$ and $H'_{ab}$ as
\begin{equation}
M = \frac{\sum_{a,b} H_{ab} H'_{ab} }{ \left(\sum_{a,b} H_{ab} H_{ab} \sum_{a,b} H'_{ab} H'_{ab}\right)^{1/2}} 
\end{equation}
Because the correlation matrices are dominated by their diagonal terms, the match $M$ will always be greater than zero. Since what we are really interested in are the cross-correlation terms, we can define a modified
match that excludes the diagonal contributions:
\begin{equation}
\overline{M} = \frac{\sum_{a\neq b} H_{ab} H'_{ab} }{ \left(\sum_{a\neq b} H_{ab} H_{ab} \sum_{a \neq b} H'_{ab} H'_{ab}\right)^{1/2}} 
\end{equation}
For realistic networks in which the noise varies between pulsars and with frequency, the sums in the above expression should be replaced by noise-weighted sums of the form
\begin{equation}
 \sum_{a, b} H_{ab} H'_{ab}  \rightarrow  \sum_{a, b} \int \frac{ H_{ab} H'_{ab} }{S_{a}(f) S_{b}(f)} \, df \,,
\end{equation}
where $S_{a}(f)$ is the sensitivity curve for pulsar $a$. The sensitivity curve is derived by convolving the noise spectrum with the gravitational wave response and
the timing model. Figure 10 shows a histogram of $M$ and $\overline{M}$ for randomly drawn sky locations for an equal-sensitivity 20 pulsar network. The width of the
distributions scale inversely with the effective number of pulsars in the array, which we define as
\begin{equation}
N_{\rm eff} = \frac{\sum_{a=1}^{N_p}  \int S_{a}(f)^{-1}  df }{S^{-1}_{\rm max}} \, ,
\end{equation}
where $S^{-1}_{\rm max} =  \max_{1 \leq a \leq N_p} \int  S_{a}(f)^{-1}  df$. For an equally sensitivity network $N_{\rm eff} = N_p$, while for a heterogeneous network  $N_{\rm eff} < N_p$. For example, the $N_p=36$ pulsar network used in the first IPTA mock data challenge has $N_{\rm eff} = 4.35$, which nicely explains the result from the previous section that shows that the IPTA array behaves more like an array with $N_p=5$ than with $N_p=20$. 

\begin{figure}[htp]
\includegraphics[clip=true,angle=0,width=0.48\textwidth]{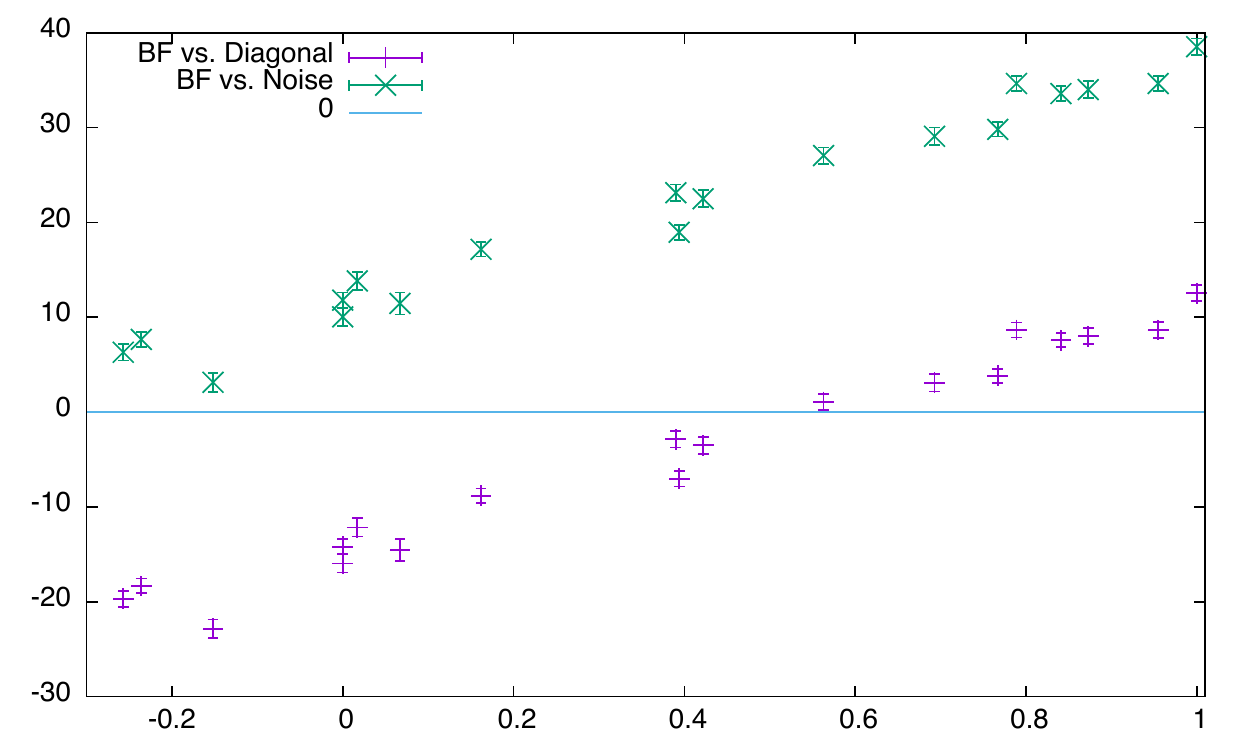} 
\caption{\label{fig:MvBF}  Log Bayes factors in favor of a detection as a function of the match $\overline{M}$ for an equal-sensitivity 20 pulsar timing array. The green points compare the
full signal model to the noise model, while the purple points compare the full signal model to a signal model with a diagonal correlation matrix.}
\end{figure}

We expect to see a correlation between the match, $\overline{M}$, and the Bayes factor between the signal and noise model. Figure 11 shows a roughly linear relationship between $\overline{M}$ and $\ln({\rm BF})$.
The simulations were for an equal-sensitivity 20 pulsar network for two noise levels using one realization of the full black hole population mode. Because random scrambles rarely produce
matches above  $\overline{M} = 0.2$ for a 20 pulsar network, the high match examples
were found by applying small random perturbations to the true pulsar locations. We see that even scrambles with $\overline{M} < 0$ can produce Bayes factors in favor of the signal model, as the diagonal components
of $H'_{ab}$ pick up power from both the Earth-term and the Pulsar-term. Another way of saying this is that the full match, $M$, which is greater than zero for all scrambles, is the relevant quantity when comparing the signal model to
the standard noise model described by (\ref{cmatrix}). As described in the previous section, we can focus the analysis on the off-diagonal, cross-correlation pattern by adding a common red ``noise'' term as a diagonal component of $C_{ab}$ to the standard noise model - we call this the diagonal model. This results in the Bayesian equivalent of the frequentist optimal statistic defined in Ref.~\cite{2009PhRvD..79h4030A, 2014arXiv1410.8256C}. 

\begin{figure}[htp]
\includegraphics[clip=true,angle=0,width=0.48\textwidth]{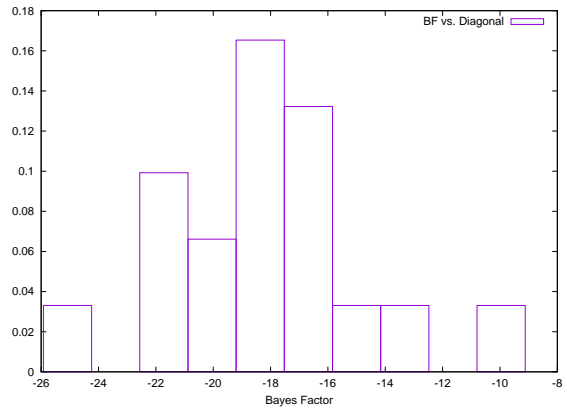} 
\caption{\label{fig:MvBF}  A histogram of the log Bayes factors between the full correlation model and a diagonal correlation model for the 18 independent sky-scrambles for an equal-sensitivity 20 pulsar timing array.
Here independent is defined as matches with the true correlation pattern and each other of  $\overline{M} < 0$. For reference, the analysis with un-scrambled sky locations gave a log Bayes factor of 12.}
\end{figure}

In the LIGO/Virgo context, thousands of independent time-slides are used to establish the false alarm rate. In the pulsar timing case we are unable to generate nearly so many independent sky-scrambles, at least
when using $\overline{M}$ as the sole measure of independence. For example, if we define two correlation patterns $H_{ab}$ to be independent if their match $\overline{M} < 0$, then the
number of independent scrambles is roughly equal to $N_{\rm eff}$. This number is derived from numerical experiments in which successive skies were generated randomly and added to the group of independent skies if
they were independent of all the other skies in the group. If the criteria for independence is relaxed to, say, $\overline{M} < 0.1$, then the number of independent skies grows by an order of magnitude. Figure 12
shows the distribution of log Bayes factors between the full and diagonal signal models for 18 mutually independent scrambles (defined as having $\overline{M} < 0$), using the same data set as Figure 11.
Notice that none of the analyses with scrambled sky locations give Bayes in favor of a gravitational wave signal with a tensor correlation pattern being present, while the analysis using the correct
sky locations gave overwhelming evidence for such a signal. With just 10 or 20 independent scrambles it is not possible to make very interesting statements about ``false alarm rates'', which in
any case are rather meaningless in the context of measuring a single realization of a signal. Rather, the sky scrambles, along with analyses of simulated signals, can be used as a way of testing
the models being used in the Bayesian analysis. It may also be that using the match to define independence significantly underestimate the number of independent sky scrambles. For example, there are
many scrambles that give the same match value $\overline{M}$ to the original array, but have very different collections of pulsar pairs in the positive and negative sectors of the Hellings-Downs curve.

\section{Summary}
\label{Sec:summary}
The detection of a stochastic gravitational wave background with PTAs requires the detection of cross-correlations between the timing residuals in multiple pulsars. When either the gravitational wave signal or the pulsar array is spatially isotropic, the values of these cross-correlations are uniquely determined by the tensor nature of the radiation via the Hellings-Downs curve. Standard pipelines for analyzing the stochastic background in PTA data assume an isotropic background when searching for these cross-correlations. We have shown that, despite the fact that realistic gravitational wave skies can contain a large degree of anisotropy, the isotropic search is remarkably robust, and only leads to loss of detection efficiency in extreme cases. This robustness being established, we have shown that we can break the expected correlations by scrambling the location of the pulsars in the sky, and that these sky scrambles result in a lower evidence for the signal model. Thus we can build confidence in the detection of a stochastic background of gravitational waves be establishing that the evidence for such a detection shrinks as the correlation matrix for the pulsars in the array is made more and more dissimilar to the Hellings-Downs values.

\acknowledgments

We benefited from discussion with participants in the Aspen Center for Physics summer program on ``Computation, systematics, and
inference for pulsar-timing arrays, and beyond'', including Steven Taylor, Jonathan Gair, Stanislav Babak, Alberto Sesana, Xavier Siemens, Sean McWilliams, Justin Ellis, Rutger van Haasteren
Joe Romano and Chiara Mingarelli. The Aspen Center for Physics is supported by National Science Foundation grant PHY-106629.
NJC was supported by NSF Physics Frontiers Center Award PFC-1430284. LS was supported by Nicol\'{a}s Yunes'  NSF CAREER Award PHY-1250636. 


\bibliography{master}

\begin{thebibliography}{38}
\expandafter\ifx\csname natexlab\endcsname\relax\def\natexlab#1{#1}\fi
\expandafter\ifx\csname bibnamefont\endcsname\relax
  \def\bibnamefont#1{#1}\fi
\expandafter\ifx\csname bibfnamefont\endcsname\relax
  \def\bibfnamefont#1{#1}\fi
\expandafter\ifx\csname citenamefont\endcsname\relax
  \def\citenamefont#1{#1}\fi
\expandafter\ifx\csname url\endcsname\relax
  \def\url#1{\texttt{#1}}\fi
\expandafter\ifx\csname urlprefix\endcsname\relax\def\urlprefix{URL }\fi
\providecommand{\bibinfo}[2]{#2}
\providecommand{\eprint}[2][]{\url{#2}}

\bibitem[{\citenamefont{Shannon et~al.}(2015)\citenamefont{Shannon, Ravi,
  Lentati, Lasky, Hobbs, Kerr, Manchester, Coles, Levin, Bailes
  et~al.}}]{Shannon25092015}
\bibinfo{author}{\bibfnamefont{R.~M.} \bibnamefont{Shannon}},
  \bibinfo{author}{\bibfnamefont{V.}~\bibnamefont{Ravi}},
  \bibinfo{author}{\bibfnamefont{L.~T.} \bibnamefont{Lentati}},
  \bibinfo{author}{\bibfnamefont{P.~D.} \bibnamefont{Lasky}},
  \bibinfo{author}{\bibfnamefont{G.}~\bibnamefont{Hobbs}},
  \bibinfo{author}{\bibfnamefont{M.}~\bibnamefont{Kerr}},
  \bibinfo{author}{\bibfnamefont{R.~N.} \bibnamefont{Manchester}},
  \bibinfo{author}{\bibfnamefont{W.~A.} \bibnamefont{Coles}},
  \bibinfo{author}{\bibfnamefont{Y.}~\bibnamefont{Levin}},
  \bibinfo{author}{\bibfnamefont{M.}~\bibnamefont{Bailes}},
  \bibnamefont{et~al.}, \bibinfo{journal}{Science}
  \textbf{\bibinfo{volume}{349}}, \bibinfo{pages}{1522} (\bibinfo{year}{2015}),
  \eprint{http://www.sciencemag.org/content/349/6255/1522.full.pdf},
  \urlprefix\url{http://www.sciencemag.org/content/349/6255/1522.abstract}.

\bibitem[{\citenamefont{{Lentati} et~al.}(2015)\citenamefont{{Lentati},
  {Taylor}, {Mingarelli}, {Sesana}, {Sanidas}, {Vecchio}, {Caballero}, {Lee},
  {van Haasteren}, {Babak} et~al.}}]{2015MNRAS.453.2576L}
\bibinfo{author}{\bibfnamefont{L.}~\bibnamefont{{Lentati}}},
  \bibinfo{author}{\bibfnamefont{S.~R.} \bibnamefont{{Taylor}}},
  \bibinfo{author}{\bibfnamefont{C.~M.~F.} \bibnamefont{{Mingarelli}}},
  \bibinfo{author}{\bibfnamefont{A.}~\bibnamefont{{Sesana}}},
  \bibinfo{author}{\bibfnamefont{S.~A.} \bibnamefont{{Sanidas}}},
  \bibinfo{author}{\bibfnamefont{A.}~\bibnamefont{{Vecchio}}},
  \bibinfo{author}{\bibfnamefont{R.~N.} \bibnamefont{{Caballero}}},
  \bibinfo{author}{\bibfnamefont{K.~J.} \bibnamefont{{Lee}}},
  \bibinfo{author}{\bibfnamefont{R.}~\bibnamefont{{van Haasteren}}},
  \bibinfo{author}{\bibfnamefont{S.}~\bibnamefont{{Babak}}},
  \bibnamefont{et~al.}, \bibinfo{journal}{\mnras}
  \textbf{\bibinfo{volume}{453}}, \bibinfo{pages}{2576} (\bibinfo{year}{2015}),
  \eprint{1504.03692}.

\bibitem[{\citenamefont{{Arzoumanian} et~al.}(2015)\citenamefont{{Arzoumanian},
  {Brazier}, {Burke-Spolaor}, {Chamberlin}, {Chatterjee}, {Christy}, {Cordes},
  {Cornish}, {Demorest}, {Deng} et~al.}}]{2015arXiv150803024A}
\bibinfo{author}{\bibfnamefont{Z.}~\bibnamefont{{Arzoumanian}}},
  \bibinfo{author}{\bibfnamefont{A.}~\bibnamefont{{Brazier}}},
  \bibinfo{author}{\bibfnamefont{S.}~\bibnamefont{{Burke-Spolaor}}},
  \bibinfo{author}{\bibfnamefont{S.}~\bibnamefont{{Chamberlin}}},
  \bibinfo{author}{\bibfnamefont{S.}~\bibnamefont{{Chatterjee}}},
  \bibinfo{author}{\bibfnamefont{B.}~\bibnamefont{{Christy}}},
  \bibinfo{author}{\bibfnamefont{J.}~\bibnamefont{{Cordes}}},
  \bibinfo{author}{\bibfnamefont{N.}~\bibnamefont{{Cornish}}},
  \bibinfo{author}{\bibfnamefont{P.}~\bibnamefont{{Demorest}}},
  \bibinfo{author}{\bibfnamefont{X.}~\bibnamefont{{Deng}}},
  \bibnamefont{et~al.}, \bibinfo{journal}{ArXiv e-prints}
  (\bibinfo{year}{2015}), \eprint{1508.03024}.

\bibitem[{\citenamefont{Siemens et~al.}(2013)\citenamefont{Siemens, Ellis,
  Jenet, and Romano}}]{Siemens:2013zla}
\bibinfo{author}{\bibfnamefont{X.}~\bibnamefont{Siemens}},
  \bibinfo{author}{\bibfnamefont{J.}~\bibnamefont{Ellis}},
  \bibinfo{author}{\bibfnamefont{F.}~\bibnamefont{Jenet}}, \bibnamefont{and}
  \bibinfo{author}{\bibfnamefont{J.~D.} \bibnamefont{Romano}},
  \bibinfo{journal}{Class.Quant.Grav.} \textbf{\bibinfo{volume}{30}},
  \bibinfo{pages}{224015} (\bibinfo{year}{2013}), \eprint{1305.3196}.

\bibitem[{\citenamefont{Hellings and Downs}(1983)}]{HellingsDowns}
\bibinfo{author}{\bibfnamefont{R.~W.} \bibnamefont{Hellings}} \bibnamefont{and}
  \bibinfo{author}{\bibfnamefont{G.~S.} \bibnamefont{Downs}},
  \bibinfo{journal}{Astrophysical Journal - Letters}  (\bibinfo{year}{1983}).

\bibitem[{\citenamefont{Cornish and Sesana}(2013)}]{Cornish:2013aba}
\bibinfo{author}{\bibfnamefont{N.~J.} \bibnamefont{Cornish}} \bibnamefont{and}
  \bibinfo{author}{\bibfnamefont{A.}~\bibnamefont{Sesana}},
  \bibinfo{journal}{Class.Quant.Grav.} \textbf{\bibinfo{volume}{30}},
  \bibinfo{pages}{224005} (\bibinfo{year}{2013}), \eprint{1305.0326}.

\bibitem[{\citenamefont{Cordes and Shannon}(2012)}]{Cordes:2011vg}
\bibinfo{author}{\bibfnamefont{J.~M.} \bibnamefont{Cordes}} \bibnamefont{and}
  \bibinfo{author}{\bibfnamefont{R.~M.} \bibnamefont{Shannon}},
  \bibinfo{journal}{Astrophys. J.} \textbf{\bibinfo{volume}{750}},
  \bibinfo{pages}{89} (\bibinfo{year}{2012}), \eprint{1106.4047}.

\bibitem[{\citenamefont{{Yardley} et~al.}(2011)\citenamefont{{Yardley},
  {Coles}, {Hobbs}, {Verbiest}, {Manchester}, {van Straten}, {Jenet}, {Bailes},
  {Bhat}, {Burke-Spolaor} et~al.}}]{2011MNRAS.414.1777Y}
\bibinfo{author}{\bibfnamefont{D.~R.~B.} \bibnamefont{{Yardley}}},
  \bibinfo{author}{\bibfnamefont{W.~A.} \bibnamefont{{Coles}}},
  \bibinfo{author}{\bibfnamefont{G.~B.} \bibnamefont{{Hobbs}}},
  \bibinfo{author}{\bibfnamefont{J.~P.~W.} \bibnamefont{{Verbiest}}},
  \bibinfo{author}{\bibfnamefont{R.~N.} \bibnamefont{{Manchester}}},
  \bibinfo{author}{\bibfnamefont{W.}~\bibnamefont{{van Straten}}},
  \bibinfo{author}{\bibfnamefont{F.~A.} \bibnamefont{{Jenet}}},
  \bibinfo{author}{\bibfnamefont{M.}~\bibnamefont{{Bailes}}},
  \bibinfo{author}{\bibfnamefont{N.~D.~R.} \bibnamefont{{Bhat}}},
  \bibinfo{author}{\bibfnamefont{S.}~\bibnamefont{{Burke-Spolaor}}},
  \bibnamefont{et~al.}, \bibinfo{journal}{\mnras}
  \textbf{\bibinfo{volume}{414}}, \bibinfo{pages}{1777} (\bibinfo{year}{2011}),
  \eprint{1102.2230}.

\bibitem[{\citenamefont{{Anholm} et~al.}(2009)\citenamefont{{Anholm},
  {Ballmer}, {Creighton}, {Price}, and {Siemens}}}]{2009PhRvD..79h4030A}
\bibinfo{author}{\bibfnamefont{M.}~\bibnamefont{{Anholm}}},
  \bibinfo{author}{\bibfnamefont{S.}~\bibnamefont{{Ballmer}}},
  \bibinfo{author}{\bibfnamefont{J.~D.~E.} \bibnamefont{{Creighton}}},
  \bibinfo{author}{\bibfnamefont{L.~R.} \bibnamefont{{Price}}},
  \bibnamefont{and}
  \bibinfo{author}{\bibfnamefont{X.}~\bibnamefont{{Siemens}}},
  \bibinfo{journal}{\prd} \textbf{\bibinfo{volume}{79}}, \bibinfo{eid}{084030}
  (\bibinfo{year}{2009}), \eprint{0809.0701}.

\bibitem[{\citenamefont{{Chamberlin} et~al.}(2015)\citenamefont{{Chamberlin},
  {Creighton}, {Siemens}, {Demorest}, {Ellis}, {Price}, and
  {Romano}}}]{2015PhRvD..91d4048C}
\bibinfo{author}{\bibfnamefont{S.~J.} \bibnamefont{{Chamberlin}}},
  \bibinfo{author}{\bibfnamefont{J.~D.~E.} \bibnamefont{{Creighton}}},
  \bibinfo{author}{\bibfnamefont{X.}~\bibnamefont{{Siemens}}},
  \bibinfo{author}{\bibfnamefont{P.}~\bibnamefont{{Demorest}}},
  \bibinfo{author}{\bibfnamefont{J.}~\bibnamefont{{Ellis}}},
  \bibinfo{author}{\bibfnamefont{L.~R.} \bibnamefont{{Price}}},
  \bibnamefont{and} \bibinfo{author}{\bibfnamefont{J.~D.}
  \bibnamefont{{Romano}}}, \bibinfo{journal}{\prd}
  \textbf{\bibinfo{volume}{91}}, \bibinfo{eid}{044048} (\bibinfo{year}{2015}),
  \eprint{1410.8256}.

\bibitem[{\citenamefont{{Finn} et~al.}(2009)\citenamefont{{Finn}, {Larson}, and
  {Romano}}}]{2009PhRvD..79f2003F}
\bibinfo{author}{\bibfnamefont{L.~S.} \bibnamefont{{Finn}}},
  \bibinfo{author}{\bibfnamefont{S.~L.} \bibnamefont{{Larson}}},
  \bibnamefont{and} \bibinfo{author}{\bibfnamefont{J.~D.}
  \bibnamefont{{Romano}}}, \bibinfo{journal}{\prd}
  \textbf{\bibinfo{volume}{79}}, \bibinfo{eid}{062003} (\bibinfo{year}{2009}),
  \eprint{0811.3582}.

\bibitem[{\citenamefont{{Chamberlin} et~al.}(2014)\citenamefont{{Chamberlin},
  {Creighton}, {Demorest}, {Ellis}, {Price}, {Romano}, and
  {Siemens}}}]{2014arXiv1410.8256C}
\bibinfo{author}{\bibfnamefont{S.~J.} \bibnamefont{{Chamberlin}}},
  \bibinfo{author}{\bibfnamefont{J.~D.~E.} \bibnamefont{{Creighton}}},
  \bibinfo{author}{\bibfnamefont{P.~B.} \bibnamefont{{Demorest}}},
  \bibinfo{author}{\bibfnamefont{J.}~\bibnamefont{{Ellis}}},
  \bibinfo{author}{\bibfnamefont{L.~R.} \bibnamefont{{Price}}},
  \bibinfo{author}{\bibfnamefont{J.~D.} \bibnamefont{{Romano}}},
  \bibnamefont{and}
  \bibinfo{author}{\bibfnamefont{X.}~\bibnamefont{{Siemens}}},
  \bibinfo{journal}{ArXiv e-prints}  (\bibinfo{year}{2014}),
  \eprint{1410.8256}.

\bibitem[{\citenamefont{{Wyithe} and {Loeb}}(2003)}]{2003ApJ...590..691W}
\bibinfo{author}{\bibfnamefont{J.~S.~B.} \bibnamefont{{Wyithe}}}
  \bibnamefont{and} \bibinfo{author}{\bibfnamefont{A.}~\bibnamefont{{Loeb}}},
  \bibinfo{journal}{\apj} \textbf{\bibinfo{volume}{590}}, \bibinfo{pages}{691}
  (\bibinfo{year}{2003}), \eprint{astro-ph/0211556}.

\bibitem[{\citenamefont{{Jaffe} and {Backer}}(2003)}]{2003ApJ...583..616J}
\bibinfo{author}{\bibfnamefont{A.~H.} \bibnamefont{{Jaffe}}} \bibnamefont{and}
  \bibinfo{author}{\bibfnamefont{D.~C.} \bibnamefont{{Backer}}},
  \bibinfo{journal}{\apj} \textbf{\bibinfo{volume}{583}}, \bibinfo{pages}{616}
  (\bibinfo{year}{2003}), \eprint{astro-ph/0210148}.

\bibitem[{\citenamefont{{Rajagopal} and {Romani}}(1995)}]{1995ApJ...446..543R}
\bibinfo{author}{\bibfnamefont{M.}~\bibnamefont{{Rajagopal}}} \bibnamefont{and}
  \bibinfo{author}{\bibfnamefont{R.~W.} \bibnamefont{{Romani}}},
  \bibinfo{journal}{\apj} \textbf{\bibinfo{volume}{446}}, \bibinfo{pages}{543}
  (\bibinfo{year}{1995}), \eprint{astro-ph/9412038}.

\bibitem[{\citenamefont{{Rosado} and {Sesana}}(2014)}]{2014MNRAS.439.3986R}
\bibinfo{author}{\bibfnamefont{P.~A.} \bibnamefont{{Rosado}}} \bibnamefont{and}
  \bibinfo{author}{\bibfnamefont{A.}~\bibnamefont{{Sesana}}},
  \bibinfo{journal}{\mnras} \textbf{\bibinfo{volume}{439}},
  \bibinfo{pages}{3986} (\bibinfo{year}{2014}), \eprint{1311.0883}.

\bibitem[{\citenamefont{{Ravi} et~al.}(2012)\citenamefont{{Ravi}, {Wyithe},
  {Hobbs}, {Shannon}, {Manchester}, {Yardley}, and
  {Keith}}}]{2012ApJ...761...84R}
\bibinfo{author}{\bibfnamefont{V.}~\bibnamefont{{Ravi}}},
  \bibinfo{author}{\bibfnamefont{J.~S.~B.} \bibnamefont{{Wyithe}}},
  \bibinfo{author}{\bibfnamefont{G.}~\bibnamefont{{Hobbs}}},
  \bibinfo{author}{\bibfnamefont{R.~M.} \bibnamefont{{Shannon}}},
  \bibinfo{author}{\bibfnamefont{R.~N.} \bibnamefont{{Manchester}}},
  \bibinfo{author}{\bibfnamefont{D.~R.~B.} \bibnamefont{{Yardley}}},
  \bibnamefont{and} \bibinfo{author}{\bibfnamefont{M.~J.}
  \bibnamefont{{Keith}}}, \bibinfo{journal}{\apj}
  \textbf{\bibinfo{volume}{761}}, \bibinfo{eid}{84} (\bibinfo{year}{2012}),
  \eprint{1210.3854}.

\bibitem[{\citenamefont{{Mingarelli} et~al.}(2013)\citenamefont{{Mingarelli},
  {Sidery}, {Mandel}, and {Vecchio}}}]{2013PhRvD..88f2005M}
\bibinfo{author}{\bibfnamefont{C.~M.~F.} \bibnamefont{{Mingarelli}}},
  \bibinfo{author}{\bibfnamefont{T.}~\bibnamefont{{Sidery}}},
  \bibinfo{author}{\bibfnamefont{I.}~\bibnamefont{{Mandel}}}, \bibnamefont{and}
  \bibinfo{author}{\bibfnamefont{A.}~\bibnamefont{{Vecchio}}},
  \bibinfo{journal}{\prd} \textbf{\bibinfo{volume}{88}}, \bibinfo{eid}{062005}
  (\bibinfo{year}{2013}), \eprint{1306.5394}.

\bibitem[{\citenamefont{{Taylor} and {Gair}}(2013)}]{2013PhRvD..88h4001T}
\bibinfo{author}{\bibfnamefont{S.~R.} \bibnamefont{{Taylor}}} \bibnamefont{and}
  \bibinfo{author}{\bibfnamefont{J.~R.} \bibnamefont{{Gair}}},
  \bibinfo{journal}{\prd} \textbf{\bibinfo{volume}{88}}, \bibinfo{eid}{084001}
  (\bibinfo{year}{2013}), \eprint{1306.5395}.

\bibitem[{\citenamefont{{Taylor} et~al.}(2015)\citenamefont{{Taylor},
  {Mingarelli}, {Gair}, {Sesana}, {Theureau}, {Babak}, {Bassa}, {Brem},
  {Burgay}, {Caballero} et~al.}}]{2015PhRvL.115d1101T}
\bibinfo{author}{\bibfnamefont{S.~R.} \bibnamefont{{Taylor}}},
  \bibinfo{author}{\bibfnamefont{C.~M.~F.} \bibnamefont{{Mingarelli}}},
  \bibinfo{author}{\bibfnamefont{J.~R.} \bibnamefont{{Gair}}},
  \bibinfo{author}{\bibfnamefont{A.}~\bibnamefont{{Sesana}}},
  \bibinfo{author}{\bibfnamefont{G.}~\bibnamefont{{Theureau}}},
  \bibinfo{author}{\bibfnamefont{S.}~\bibnamefont{{Babak}}},
  \bibinfo{author}{\bibfnamefont{C.~G.} \bibnamefont{{Bassa}}},
  \bibinfo{author}{\bibfnamefont{P.}~\bibnamefont{{Brem}}},
  \bibinfo{author}{\bibfnamefont{M.}~\bibnamefont{{Burgay}}},
  \bibinfo{author}{\bibfnamefont{R.~N.} \bibnamefont{{Caballero}}},
  \bibnamefont{et~al.}, \bibinfo{journal}{Physical Review Letters}
  \textbf{\bibinfo{volume}{115}}, \bibinfo{eid}{041101} (\bibinfo{year}{2015}),
  \eprint{1506.08817}.

\bibitem[{\citenamefont{{Cornish} and {van
  Haasteren}}(2014)}]{2014arXiv1406.4511C}
\bibinfo{author}{\bibfnamefont{N.~J.} \bibnamefont{{Cornish}}}
  \bibnamefont{and} \bibinfo{author}{\bibfnamefont{R.}~\bibnamefont{{van
  Haasteren}}}, \bibinfo{journal}{ArXiv e-prints}  (\bibinfo{year}{2014}),
  \eprint{1406.4511}.

\bibitem[{\citenamefont{Sampson et~al.}(2015)\citenamefont{Sampson, Cornish,
  and McWilliams}}]{Sampson:2015ada}
\bibinfo{author}{\bibfnamefont{L.}~\bibnamefont{Sampson}},
  \bibinfo{author}{\bibfnamefont{N.~J.} \bibnamefont{Cornish}},
  \bibnamefont{and} \bibinfo{author}{\bibfnamefont{S.~T.}
  \bibnamefont{McWilliams}}, \bibinfo{journal}{Phys.Rev.}
  \textbf{\bibinfo{volume}{D91}}, \bibinfo{pages}{084055}
  (\bibinfo{year}{2015}), \eprint{1503.02662}.

\bibitem[{\citenamefont{{Goggans} and {Chi}}(2004)}]{2004AIPC..707...59G}
\bibinfo{author}{\bibfnamefont{P.~M.} \bibnamefont{{Goggans}}}
  \bibnamefont{and} \bibinfo{author}{\bibfnamefont{Y.}~\bibnamefont{{Chi}}}, in
  \emph{\bibinfo{booktitle}{Bayesian Inference and Maximum Entropy Methods in
  Science and Engineering}}, edited by \bibinfo{editor}{\bibfnamefont{G.~J.}
  \bibnamefont{{Erickson}}} \bibnamefont{and}
  \bibinfo{editor}{\bibfnamefont{Y.}~\bibnamefont{{Zhai}}}
  (\bibinfo{year}{2004}), vol. \bibinfo{volume}{707} of
  \emph{\bibinfo{series}{American Institute of Physics Conference Series}}, pp.
  \bibinfo{pages}{59--66}.

\bibitem[{\citenamefont{{Swendsen} and {Wang}}(1986)}]{1986PhRvL..57.2607S}
\bibinfo{author}{\bibfnamefont{R.~H.} \bibnamefont{{Swendsen}}}
  \bibnamefont{and} \bibinfo{author}{\bibfnamefont{J.-S.}
  \bibnamefont{{Wang}}}, \bibinfo{journal}{Physical Review Letters}
  \textbf{\bibinfo{volume}{57}}, \bibinfo{pages}{2607} (\bibinfo{year}{1986}).

\bibitem[{\citenamefont{Taylor and Gair}(2013)}]{Taylor:2013esa}
\bibinfo{author}{\bibfnamefont{S.~R.} \bibnamefont{Taylor}} \bibnamefont{and}
  \bibinfo{author}{\bibfnamefont{J.~R.} \bibnamefont{Gair}},
  \bibinfo{journal}{Phys. Rev.} \textbf{\bibinfo{volume}{D88}},
  \bibinfo{pages}{084001} (\bibinfo{year}{2013}), \eprint{1306.5395}.

\bibitem[{\citenamefont{Gair et~al.}(2014)\citenamefont{Gair, Romano, Taylor,
  and Mingarelli}}]{Gair:2014rwa}
\bibinfo{author}{\bibfnamefont{J.}~\bibnamefont{Gair}},
  \bibinfo{author}{\bibfnamefont{J.~D.} \bibnamefont{Romano}},
  \bibinfo{author}{\bibfnamefont{S.}~\bibnamefont{Taylor}}, \bibnamefont{and}
  \bibinfo{author}{\bibfnamefont{C.~M.} \bibnamefont{Mingarelli}},
  \bibinfo{journal}{Phys. Rev.} \textbf{\bibinfo{volume}{D90}},
  \bibinfo{pages}{082001} (\bibinfo{year}{2014}), \eprint{1406.4664}.

\bibitem[{\citenamefont{Cornish and van Haasteren}(2014)}]{Cornish:2014rva}
\bibinfo{author}{\bibfnamefont{N.~J.} \bibnamefont{Cornish}} \bibnamefont{and}
  \bibinfo{author}{\bibfnamefont{R.}~\bibnamefont{van Haasteren}}
  (\bibinfo{year}{2014}), \eprint{1406.4511}.

\bibitem[{\citenamefont{Shannon and Cordes}(2010)}]{Shannon:2010bv}
\bibinfo{author}{\bibfnamefont{R.~M.} \bibnamefont{Shannon}} \bibnamefont{and}
  \bibinfo{author}{\bibfnamefont{J.~M.} \bibnamefont{Cordes}},
  \bibinfo{journal}{Astrophys. J.} \textbf{\bibinfo{volume}{725}},
  \bibinfo{pages}{1607} (\bibinfo{year}{2010}), \eprint{1010.4794}.

\bibitem[{\citenamefont{Taylor et~al.}(2013)\citenamefont{Taylor, Gair, and
  Lentati}}]{Taylor:2012wv}
\bibinfo{author}{\bibfnamefont{S.~R.} \bibnamefont{Taylor}},
  \bibinfo{author}{\bibfnamefont{J.~R.} \bibnamefont{Gair}}, \bibnamefont{and}
  \bibinfo{author}{\bibfnamefont{L.}~\bibnamefont{Lentati}},
  \bibinfo{journal}{Phys. Rev.} \textbf{\bibinfo{volume}{D87}},
  \bibinfo{pages}{044035} (\bibinfo{year}{2013}), \eprint{1210.6014}.

\bibitem[{\citenamefont{Lentati et~al.}(2013)\citenamefont{Lentati, Alexander,
  Hobson, Taylor, Gair, Balan, and van Haasteren}}]{Lentati:2012xb}
\bibinfo{author}{\bibfnamefont{L.}~\bibnamefont{Lentati}},
  \bibinfo{author}{\bibfnamefont{P.}~\bibnamefont{Alexander}},
  \bibinfo{author}{\bibfnamefont{M.~P.} \bibnamefont{Hobson}},
  \bibinfo{author}{\bibfnamefont{S.}~\bibnamefont{Taylor}},
  \bibinfo{author}{\bibfnamefont{J.}~\bibnamefont{Gair}},
  \bibinfo{author}{\bibfnamefont{S.~T.} \bibnamefont{Balan}}, \bibnamefont{and}
  \bibinfo{author}{\bibfnamefont{R.}~\bibnamefont{van Haasteren}},
  \bibinfo{journal}{Phys. Rev.} \textbf{\bibinfo{volume}{D87}},
  \bibinfo{pages}{104021} (\bibinfo{year}{2013}), \eprint{1210.3578}.

\bibitem[{\citenamefont{Ellis et~al.}(2013)\citenamefont{Ellis, Siemens, and
  van Haasteren}}]{Ellis:2013nrb}
\bibinfo{author}{\bibfnamefont{J.~A.} \bibnamefont{Ellis}},
  \bibinfo{author}{\bibfnamefont{X.}~\bibnamefont{Siemens}}, \bibnamefont{and}
  \bibinfo{author}{\bibfnamefont{R.}~\bibnamefont{van Haasteren}},
  \bibinfo{journal}{Astrophys. J.} \textbf{\bibinfo{volume}{769}},
  \bibinfo{pages}{63} (\bibinfo{year}{2013}), \eprint{1302.1903}.

\bibitem[{\citenamefont{Aasi et~al.}(2015)}]{TheLIGOScientific:2014jea}
\bibinfo{author}{\bibfnamefont{J.}~\bibnamefont{Aasi}} \bibnamefont{et~al.}
  (\bibinfo{collaboration}{LIGO Scientific}),
  \bibinfo{journal}{Class.Quant.Grav.} \textbf{\bibinfo{volume}{32}},
  \bibinfo{pages}{074001} (\bibinfo{year}{2015}), \eprint{1411.4547}.

\bibitem[{\citenamefont{Accadia et~al.}(2015)\citenamefont{Accadia, Agathos,
  Allocca, Astone, Ballardin et~al.}}]{Accadia:2015pda}
\bibinfo{author}{\bibfnamefont{T.}~\bibnamefont{Accadia}},
  \bibinfo{author}{\bibfnamefont{M.}~\bibnamefont{Agathos}},
  \bibinfo{author}{\bibfnamefont{A.}~\bibnamefont{Allocca}},
  \bibinfo{author}{\bibfnamefont{P.}~\bibnamefont{Astone}},
  \bibinfo{author}{\bibfnamefont{G.}~\bibnamefont{Ballardin}},
  \bibnamefont{et~al.}, pp. \bibinfo{pages}{261--270} (\bibinfo{year}{2015}).

\bibitem[{\citenamefont{Aasi et~al.}(2012)}]{Aasi:2012wd}
\bibinfo{author}{\bibfnamefont{J.}~\bibnamefont{Aasi}} \bibnamefont{et~al.}
  (\bibinfo{collaboration}{VIRGO}), \bibinfo{journal}{Class.Quant.Grav.}
  \textbf{\bibinfo{volume}{29}}, \bibinfo{pages}{155002}
  (\bibinfo{year}{2012}), \eprint{1203.5613}.

\bibitem[{\citenamefont{Cornish and Littenberg}(2015)}]{Cornish:2014kda}
\bibinfo{author}{\bibfnamefont{N.~J.} \bibnamefont{Cornish}} \bibnamefont{and}
  \bibinfo{author}{\bibfnamefont{T.~B.} \bibnamefont{Littenberg}},
  \bibinfo{journal}{Class.Quant.Grav.} \textbf{\bibinfo{volume}{32}},
  \bibinfo{pages}{135012} (\bibinfo{year}{2015}), \eprint{1410.3835}.

\bibitem[{\citenamefont{Littenberg and Cornish}(2015)}]{Littenberg:2014oda}
\bibinfo{author}{\bibfnamefont{T.~B.} \bibnamefont{Littenberg}}
  \bibnamefont{and} \bibinfo{author}{\bibfnamefont{N.~J.}
  \bibnamefont{Cornish}}, \bibinfo{journal}{Phys.Rev.}
  \textbf{\bibinfo{volume}{D91}}, \bibinfo{pages}{084034}
  (\bibinfo{year}{2015}), \eprint{1410.3852}.

\bibitem[{\citenamefont{Blackburn et~al.}(2008)\citenamefont{Blackburn,
  Cadonati, Caride, Caudill, Chatterji et~al.}}]{Blackburn:2008ah}
\bibinfo{author}{\bibfnamefont{L.}~\bibnamefont{Blackburn}},
  \bibinfo{author}{\bibfnamefont{L.}~\bibnamefont{Cadonati}},
  \bibinfo{author}{\bibfnamefont{S.}~\bibnamefont{Caride}},
  \bibinfo{author}{\bibfnamefont{S.}~\bibnamefont{Caudill}},
  \bibinfo{author}{\bibfnamefont{S.}~\bibnamefont{Chatterji}},
  \bibnamefont{et~al.}, \bibinfo{journal}{Class.Quant.Grav.}
  \textbf{\bibinfo{volume}{25}}, \bibinfo{pages}{184004}
  (\bibinfo{year}{2008}), \eprint{0804.0800}.

\bibitem[{\citenamefont{Was et~al.}(2010)\citenamefont{Was, Bizouard, Brisson,
  Cavalier, Davier et~al.}}]{Was:2009vh}
\bibinfo{author}{\bibfnamefont{M.}~\bibnamefont{Was}},
  \bibinfo{author}{\bibfnamefont{M.-A.} \bibnamefont{Bizouard}},
  \bibinfo{author}{\bibfnamefont{V.}~\bibnamefont{Brisson}},
  \bibinfo{author}{\bibfnamefont{F.}~\bibnamefont{Cavalier}},
  \bibinfo{author}{\bibfnamefont{M.}~\bibnamefont{Davier}},
  \bibnamefont{et~al.}, \bibinfo{journal}{Class.Quant.Grav.}
  \textbf{\bibinfo{volume}{27}}, \bibinfo{pages}{015005}
  (\bibinfo{year}{2010}), \eprint{0906.2120}.

\end{thebibliography}

\end{document}